\documentclass[journal,10pt,twocolumn]{IEEEtran}

\usepackage{cite}
\usepackage{amssymb}
\usepackage{amsmath}

\usepackage{graphicx}
\usepackage{enumitem}
\usepackage{bm}
\usepackage{mathtools}

\usepackage{amsfonts}
\usepackage{stackengine}
\usepackage{xcolor}
\usepackage{stfloats}
\usepackage{array}
\newcolumntype{M}[1]{>{\centering\arraybackslash}m{#1}}
\newcolumntype{P}[1]{>{\centering\arraybackslash}p{#1}}
\stackMath
\usepackage{algorithm}
\usepackage[export]{adjustbox}
\usepackage[noend]{algpseudocode}
\usepackage{multirow} 
\usepackage{framed}
\newcommand{\mybar}[1]{\makebox[0pt]{$\phantom{#1}\overline{\phantom{#1}}$}#1}

\newcommand{\kernel}{{\ooalign{$k$\cr\raisebox{0.2em}{\kern0.08em--}\cr}
}}
\DeclareRobustCommand{\mhl}[1]{%
	\ifmmode\text{\color{red}{$#1$}}\else{\color{black}{#1}}\fi
}
\makeatletter
\newcommand{\vast}{\bBigg@{3}}
\makeatother

\makeatletter
\@ifdefinable\@latex@chi{\let\@latex@chi\chi}
\renewcommand*\chi{{\@latex@chi\smash[t]{\mathstrut}}} 
\makeatletter
\def\mathclap#1{\text{\hbox to 0pt{\hss$\mathsurround=0pt#1$\hss}}}

\makeatother
\def\mathclap#1{\text{\hbox to 0pt{\hss$\mathsurround=0pt#1$\hss}}}

\usepackage[font=footnotesize]{caption}
\usepackage[font=footnotesize]{subcaption} 

\usepackage{relsize}
\usepackage{bm}
\usepackage{soul}
\usepackage{empheq}
\usepackage{placeins}

\begin{document}

	\title{Reconciling Radio Tomographic Imaging with Phaseless Inverse Scattering}
	
	\author{Amartansh~Dubey,~\IEEEmembership{Graduate Student Member,~IEEE}, Zan~Li,~\IEEEmembership{Student Member,~IEEE},  and~Ross~Murch,~\IEEEmembership{Fellow,~IEEE}\thanks{This work was supported by the Hong Kong Research Grants Council with the Collaborative Research Fund C6012-20G. Corresponding Author:
			Amartansh Dubey.}\thanks{Amartansh Dubey and Zan Li are with the Department of Electronic and Computer Engineering, Hong Kong University of Science and Technology (HKUST), Hong Kong, (e-mail: \protect{}{adubey@connect.ust.hk; zligq@connect.ust.hk})}. \thanks{Ross Murch is with the Department of Electronic and Computer Engineering
		and the Institute of Advanced Study both at the Hong Kong University
		of Science and Technology (HKUST), Hong Kong (e-mail: eermurch@ust.hk).}}

	\maketitle

\begin{abstract}
Radio Tomographic Imaging (RTI) is a phaseless imaging approach that can provide shape reconstruction and localization of objects using received signal strength (RSS) measurements. RSS measurements can be straightforwardly obtained from wireless networks such as Wi-Fi and therefore RTI has been extensively researched and accepted as a good indoor RF imaging technique. However, RTI is formulated on empirical models using an assumption of light-of-sight (LOS) propagation that does not account for intricate scattering effects. There are two main objectives of this work. The first objective is to reconcile and compare the empirical RTI model with formal inverse scattering approaches to better understand why RTI is an effective RF imaging technique. The second objective is to obtain straightforward enhancements to RTI, based on inverse scattering, to enhance its performance. The resulting enhancements can provide reconstructions of the shape and also material properties of the objects that can aid image classification. We also provide numerical and experimental results to compare RTI with the enhanced RTI for indoor imaging applications using low-cost 2.4 GHz Wi-Fi transceivers. These results show that the enhanced RTI can outperform RTI while having similar computational complexity to RTI.
\end{abstract}


\section{Introduction}
\label{Sec_Intro}
Wireless communication infrastructure such as Wi-Fi and cellular systems can now be found in nearly any part of the populated world. This near-universal coverage of wireless communication frequency (RF) signals has motivated the development of services, other than communications, that leverage this infrastructure. Examples of these services include localization, tracking, RF imaging, and RF energy harvesting. One advantage of leveraging the use of RF signals is that they can travel through objects, providing a non-intrusive imaging approach that can see through obstacles such as walls. This is not possible with other imaging technologies such as infrared, Laser, and visible light {\cite{depatla2015x, 6881182, 1545232}}.

Wireless-based localization and imaging methods can be categorized into active device-based and device-free methods. The device-free methods do not require any active device to be included on the target and are therefore more convenient than device-based methods. There has been extensive research performed on device-free localization using Wi-Fi which includes passive coherent localization \cite{Pastina2015, 8396311, chetty2011through}, ``fingerprint'' methods \cite{adib2013see, adib2015multi, Guo2017, WangFIngerprint2017, shi2018accurate} and MIMO radar \cite{sun2018single, liu2020massive, zhang2015generalized, IPIN, 7426565}. These techniques can provide good localization accuracy. However, it is important to note the difference between \textit{localization} and \textit{imaging}. Localization refers to the tracking of the centroid of a moving target whereas imaging is a more intricate task that also requires the estimation of the shape, size, and properties (such as refractive index) of the target. Indoor imaging has a wide range of real-world applications including security and surveillance, smart radio home/buildings, indoor mapping and navigation, and health monitoring.

In this work, we focus on imaging techniques that can utilize phaseless (RSS) Wi-Fi measurements for indoor applications. Using phaseless-data greatly simplifies the measurement system by removing the need of any precise synchronization and calibration between multiple Wi-Fi nodes. As a result, the RSS data required for imaging can be collected using off-the-shelf Wi-Fi devices. There are only a few techniques available to perform indoor imaging using Wi-Fi RSS data. These are either based on empirical models developed by the wireless systems community or physics-based inverse scattering models developed by the electromagnetic community. In the wireless systems community, one of the common techniques is Radio Tomographic Imaging (RTI). In RTI, the target's location and shape are estimated by modeling the shadowing or attenuation caused by the target along the line-of-sight (LOS) path while ignoring the multipath effects. RTI has been extensively researched over the last decade \cite{Patwari2017, Patwari2015, Patwari2014, Patwari2010,1Patwari2013, Savazzi2014, WangRTI32016, WangRTI22017, WangRTI2016, TomicRTI2015}.

The electromagnetic community approaches the Wi-Fi RSS-based indoor imaging problem as an inverse scattering problem. In this approach, formal electromagnetic models that include intricate wave phenomena such as diffraction, scattering, and attenuation are utilized \cite{chen2018computational, bates1991manageable, Xudongchen, chen2010subspace, dubeyTGRS}. Several nonlinear and deep learning-based phaseless inverse scattering techniques have been proposed using this approach \cite{Xudongchen, chen2010subspace, chen2018computational}. However, due to their nonlinear nature, these techniques are computationally expensive and require data from precision experimental facilities. In addition, these techniques cannot handle strong scattering objects that are typically present in an indoor environment \cite{dubeyTGRS}. For these reasons, no existing state-of-the-art formal nonlinear or deep learning-based phaseless inverse scattering techniques have been used to demonstrate Wi-Fi RSS-based indoor imaging \cite{chen2018computational, Xudongchen}.

A relatively new class of imaging techniques have been proposed recently that combine the advantages of both RTI and physics-based inverse scattering techniques. These include approximate inverse scattering techniques that utilize linear phaseless formulations (similar to RTI) but also approximately account for intricate wave scattering effects \cite{dubeyTGRS, depatla2015x}. 
For example, a recently proposed inverse scattering technique known as the extended phaseless Rytov approximation for low-loss media (xPRA-LM) can be easily transformed into a linear phaseless form \cite{dubeyTGRS}. It has been shown \cite{dubeyTGRS} that xPRA-LM has a wide validity range and that if the target objects exhibit low-loss, have piecewise homogeneous distribution of permittivity, and are electrically large, the validity range of of xPRA-LM is  $\epsilon_r\le77$  whereas the validity range of state-of-the-art phaseless nonlinear methods do not currently extend to $\epsilon_r\le77$ \cite{Xudongchen, chen2018computational, chen2010subspace}. Another feature of xPRA-LM is that its formulation explicitly includes the loss component of materials. Nearly all objects in the indoor environment have a low-loss component and typical loss values are listed in \cite{dubeyTGRS, komarov2005permittivity, Productnote, 4562803}. In particular the loss tangents of typical objects, such as wood and glass, are of the order of 0.1 and this loss component should be included in formulations of indoor imaging and is another feature of xPRA-LM  \cite{dubeyTGRS}.
	
It should also be noted that all the imaging methods discussed, including RTI, xPRA-LM and non-linear phaseless inverse scattering techniques require 20 or more sensing nodes. In many indoor imaging scenarios this is not practical and therefore methods have also been proposed to reduce the number of sensing nodes. This includes utilizing frequency and antenna pattern diversity \cite{10241308, 9967760}. This work has shown that the number of sensing nodes can be reduced by around 50\% but further effort is still required to reduce the numbers further and this remains an active area of research.

In this work we reconcile RTI and xPRA-LM using theoretical, numerical and experimental comparisons. The key objective of this work is to answer two important questions. First, why RTI, a straightforward linear technique with empirical modeling provides very good experimental results.  Second, how can we enhance the performance of RTI by incorporating straightforward modifications based on formal inverse scattering techniques. Answering these questions leads to the following contributions:
\begin{enumerate}
	\item Reconciling RTI with formal inverse scattering formulations. Provide analysis that shows the empirical model used in RTI can be related to inverse scattering formulations and obtain analytical justifications for the success and limitations of RTI.
	\item Propose enhancements to RTI to improve its performance while keeping computational complexity as low as the original linear RTI model. These enhancements are derived using the xPRA-LM inverse scattering model. Unlike RTI which only provides low resolution shape reconstruction, the proposed enhanced RTI can provide accurate estimation of shape as well as the material properties such as attenuation coefficient (which is directly related to the refractive index and relative permittivity).
	\item Provide numerical and experimental results in real indoor environments with off-the-shelf 2.4 GHz Wi-Fi devices to compare performance of RTI and the enhanced RTI. 
\end{enumerate}
The work is different from previous work \cite{dubeyTGRS} in that we reconcile a well-known empirical RTI model with physics based xPRA-LM model using theoretical, numerical and experimental comparisons. The proposed xRTI technique in this work is also an extension of \cite{dubeyTGRS} and was not included in \cite{dubeyTGRS}. This work helps answer why RTI, a straightforward linear technique with empirical modeling provides very good experimental results and provides insight on how to enhance the performance of RTI by incorporating straightforward modifications based on formal inverse scattering techniques. 

\subsection{Organization and notations}
\label{Sec_Organization}
Organization and Notation: Matrices are denoted using double bars over upper-case bold letter ($\mybar{\mybar{\textbf{X}}}$) and vectors are denoted  using single bar over lower-case bold letter ($\mybar{\textbf{x}}$). Lower-case bold letters without bar represents position vector (${\textbf{x}}$) and italic letter ($x$) are used to represent scalar parameters.

\section{Problem Formulation}
\label{Sec_ProblemSetup}
\subsection{Problem setup and Preliminaries}
\begin{figure}[h]
	\centering
	\includegraphics[width=2.2in]{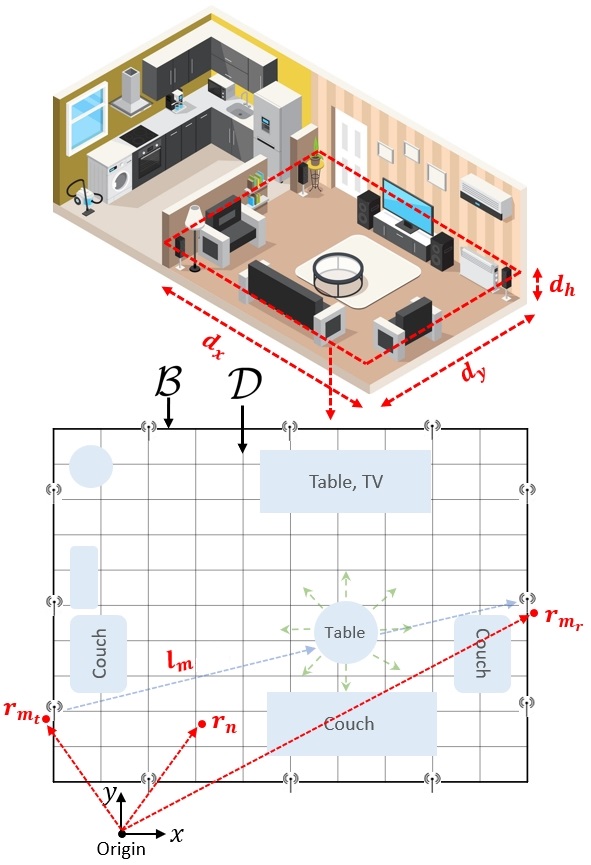}
	\caption{The DOI with wireless transceiver nodes on its boundary $ {\mathcal{B}}$. The transmitter $m_t$ and receiver $m_r$ are located at ${\bf r}_{m_t} \in {\mathcal{B}}$ and ${\bf r}_{m_r} \in {\mathcal{B}}$ respectively. The location of the $n^{th}$ grid in the DOI is denoted as ${\bf r}_n  \in {\mathcal{D}}$.}
	\label{RTInetwork2}
\end{figure}

Fig. \ref{RTInetwork2} shows a two-dimensional (2D) domain of Interest (DOI) ${\mathcal{D}}\subset \mathbb{R}^{2}$ situated in an indoor region of a building. Identical Wi-Fi transceiver nodes operating at 2.4 GHz are placed on the DOI boundary (denoted as $\mathcal{B} \subset \mathbb{R}^{2}$). There are in total $M$ transceiver nodes, each of which transmit and receive signals to acquire RSS measurements of the links between nodes. These nodes cannot transmit and receive at the same time. Therefore, the total number of unique wireless links is $L=M(M-1)/2$. $l_{m_t, m_r}$ denotes the link between the transmitting node $m_t$ (at ${\bf r}_{m_t}$) and the receiving node $m_r$ (at ${\bf r}_{m_r}$). For brevity, $l_{m_t, m_r}$ is written as $l$, where $l=1,2,...,L$. For the remainder of this work, we use the subscripts $m_t$, $m_r$ and $l$ to refer to the transmitter, receiver and corresponding wireless link respectively for all relevant quantities.

We discretize the DOI into $N = n_x \times n_y$ rectangular grids, each of size $\Delta d_x \times \Delta d_y$, where ${\bf r}_n$ denotes the location of the $n^{th}$ grid. 
The target objects inside the DOI (at any given time instant $t$) are characterized by relative permittivity $\epsilon_r(\bm{r}_n, t)=\epsilon_R(\bm{r}_n, t)+j\epsilon_I(\bm{r}_n, t)$. This parameter is also related to the complex-valued refractive index $\nu(\bm{r}_n, t)=\nu_R(\bm{r}_n, t)+j\nu_I(\bm{r}_n, t)$  where $\epsilon_r(\bm{r}_n, t)= \nu^2(\bm{r}_n, t)$. The time variations considered are assumed very much slower than the propagation time of the RF waves through the DOI.  

In the RTI community, the focus is on imaging the attenuation profile which is related to the imaginary part of refractive index, $\operatorname*{Im}\left(\nu(\bm{r}_n, t)\right) = \nu_I(\bm{r}_n, t)$. It provides an estimate of the attenuation strength whereas the real part of refractive index, $\operatorname*{Re}(\nu(\bm{r}_n, t)) = \nu_R(\bm{r}_n, t)$ can be thought of as characterizing the scattering strength. Under a low-loss assumption (which holds true for most objects around us \cite{dubeyTGRS}), the attenuation parameter \cite{hecht2012optics, bornwolf} is defined as
\begin{equation}
	\label{Eq_nu_I}
	\begin{aligned}
		\alpha(\bm{r}_n, t) = \frac{4\pi\nu_I(\bm{r}_n, t)}{\lambda_0} \approx \frac{2\pi \epsilon_I(\bm{r}_n, t)}{\lambda_0 \sqrt{\epsilon_R(\bm{r}_n, t)}}
	\end{aligned}
\end{equation}
where, $\lambda_0$ is incident wavelength which is $12.5$ cm for a 2.4 GHz Wi-Fi signals. 

We also introduce a scattering parameter, $s(\bm{r}_n, t)$, which is related to the contrast in the real part of refractive index,
\begin{equation}
	\label{Eq_nu_R}
	\begin{aligned}
		 {s(\bm{r}_n, t) = \frac{4\pi}{\lambda_0}\left(\nu_R(\bm{r}_n, t)-1\right) \approx \frac{4\pi}{\lambda_0}\left(\sqrt{\epsilon_R(\bm{r}_n, t)}-1\right)}  
	\end{aligned}
\end{equation}

The spatial DOI profile at any time instant $t$ can be characterized by $\alpha(\bm{r}_n, t)$ or $s(\bm{r}_n, t)$ and used as alternatives to using permittivity $\epsilon_r(\bm{r}_n, t)$ or complex refractive index $\nu(\bm{r}_n, t)$. In RTI and in this work, the focus is on reconstructing $\alpha(\bm{r}_n, t)$. 

In discrete form, we can express the spatial attenuation and scattering profile of the DOI at any time instant $t$ in vector form as $\mybar{\bm \alpha}(t) \in \mathcal{R}^{N\times 1} $ and $\mybar{\textbf{s}}(t) \in \mathcal{R}^{N\times 1} $ respectively, where the elements $\left[\mybar{\bm \alpha}(t)\right]_n$ and $\left[\mybar{\textbf{s}}(t)\right]_n$ represent attenuation and scattering parameters of the $n^{th}$ grid inside the DOI.

\subsection{Propagation Characterization with Background Scattering Subtraction}
The propagation loss of the $l^{th}$ link is written as $P_{l}\big(\mybar{\bm \alpha}(t),\ \mybar{\textbf{s}}(t)\big)$ (in dB) and forms our definition of RSS. The measured RSS in the $l^{th}$ link primarily depends on the node locations $({\bf r}_{m_t}, {\bf r}_{m_r})$ and the DOI's spatial profiles given by $\mybar{\bm \alpha}(t)$ and $\mybar{\textbf{s}}(t)$. Determining $P_{l}\big(\mybar{\bm \alpha}(t),\ \mybar{\textbf{s}}(t)\big)$ accurately in this scattering environment at these frequencies (around 2.4 GHz) requires full electromagnetic formulations and simulation using tools such as CST Studio Suite. RTI sidesteps these intricate formulations using two approaches: 1) dividing the propagation components into known and unknown components and 2) removing the background scattering from the RSS measurements. 

\subsubsection{Propagation Model}
In RTI, the measured RSS (in dB) is modeled in component form at time $t$ as
\begin{equation}
\label{Eq_ConRTI_0}
\begin{aligned}
	P_{l}(\mybar{\bm \alpha}(t),\mybar{\textbf{s}}(t))  = \underbrace{P^0_{l}}_{\substack{\text{ \tiny Transmitted} \\ \text{\tiny Power}}} & - \underbrace{\text{PL}_{l}}_{\substack{\text{ \tiny Path Loss,} \\ \text{\tiny }}} + \underbrace{G_{l}}_{\substack{\text{ \tiny Antenna Gains,} \\ \text{\tiny}}} + \\ & \underbrace{A_{l}(\mybar{\bm \alpha}(t))}_{\substack{\text{ \tiny Shadowing}\\ \text{ \tiny or attenuation loss}}} + \underbrace{F_{l}( \mybar{\textbf{s}}(t))}_{\substack{\text{ \tiny Scattering loss}}} + \underbrace{n^e_{l}( t)}_{\substack{\text{ \tiny experimental or} \\ \text{\tiny measurement noise} }}
\end{aligned}
\end{equation}
where, $P^0_{l}$ is the transmitted power, $\text{PL}_l$ is free-space path loss and $G_{l}$ are the combined gains of the transmitter and receiver antennas and all are assumed temporally invariant. $A_{l}$ and $F_{l}$ represent shadowing and scattering losses due to the objects inside the DOI and are functions of $\mybar{\bm \alpha}(t)$ and $\mybar{\textbf{s}}(t)$ respectively. $n^e_{l}( t)$ represents noise and modeling errors.

Accurately predicting the propagation loss $P_l$ depends on developing an accurate model for $A_l$ and $F_l$ and also knowledge of both  $\mybar{\bm \alpha}(t)$ and $\mybar{\textbf{s}}(t)$. Propagation prediction, such as this, is also formally referred to as a direct scattering problem in electromagnetic literature \cite{chen2018computational, DubeyTAP}. For example in RTI $A_l$ is modeled empirically while $F_l$ is modeled as noise and later we detail more accurate approaches. 

In contrast, from an inverse scattering or imaging perspective, we use measurements of $P_{l}$, to estimate $\mybar{\bm \alpha}(t)$ and provide an image of the DOI. Two issues however need to be addressed before the inverse problem can be solved. The first is handling the large amount of background scattering or clutter that results from stationary fixtures in the environment such as walls, ceilings, floors and furniture. The second is system calibration that obtains $P^0_{l}$, $\text{PL}_{l}$ and $G_{l}$ so they can be calibrated out from the reconstruction process. Both issues can be addressed by using temporal background propagation subtraction.  

\subsubsection{Temporal Background Propagation Subtraction}
RTI performs temporal background propagation subtraction by subtracting RSS measurements from two different time frames $t_1$ and  $t_2=t_1+\Delta t$. The resulting RSS measurements (in dB) are then found as 
\begin{equation}
	\label{Eq_NewConRTI2}
		\Delta P_{l}  = P_{l}\big(\mybar{\bm \alpha}(t_2),\ \mybar{\textbf{s}}(t_2)\big) - P_{l}\big(\mybar{\bm \alpha}(t_1),\ \mybar{\textbf{s}}(t_1)\big) 
\end{equation}
where $\Delta t$ captures changes that occur in the DOI.  $\Delta t$ can be in the range of a few seconds (such as the movement of people or placing a new object inside the DOI), a few hours (such as the movement of furniture or crowd movement) to a few days (such as architectural changes or variation in the electrical properties of existing objects). 

It can be seen from (\ref{Eq_NewConRTI2}) that the subtraction process provides straightforward calibration by removing all stationary quantities in (\ref{Eq_ConRTI_0}) such as $P^0_{l}$, $\text{PL}_{l}$ and $G_{l}$.

The subtraction process also removes background scattering or clutter that results from stationary fixtures in the environment such as walls, ceilings, floors and furniture.
The basis for using the background propagation subtraction approach is predicated on realizing that the changes in the indoor environment are usually minor compared to the background scatterers consisting of walls and furniture. In particular, the stationary background objects exhibit scattering that is often significant however the objects causing it are usually spatially separate and distinct from the moving or moved objects that are of interest. This
 implies that scattering or interaction between the moving and stationary objects will usually not dominate the wave phenomena.
 Subtracting out the stationary component will leave behind the signal predominately relating only to those
 moving objects better satisfying the assumptions of RTI and xPRA-LM. 
 
 Under (\ref{Eq_NewConRTI2}), we therefore solve two key problems highlighted previously regarding inverse scattering. 

Under (\ref{Eq_NewConRTI2}) we can image spatial changes in the DOI profile. Applications including security and surveillance, tracking people's movements, through-the-wall imaging/tracking, health monitoring, smart radio home, non-destructive evaluation of objects, are based on this approach \cite{Patwari2010, Patwari2015}.

Utilizing (\ref{Eq_NewConRTI2}), equation (\ref{Eq_ConRTI_0}) becomes
\begin{equation}
	\label{Eq_ConRTI11}
	\begin{aligned}
		\Delta P_{l} & = P_{l}\big(\mybar{\bm \alpha}(t_2),\ \mybar{\textbf{s}}(t_2)\big) - P_{l}\big(\mybar{\bm \alpha}(t_1),\ \mybar{\textbf{s}}(t_1)\big) \\
		&= - \biggl[A_{l}\big(\mybar{\bm \alpha}(t_2),\ \mybar{\textbf{s}}(t_2)\big) - A_{l}\big(\mybar{\bm \alpha}(t_1),\ \mybar{\textbf{s}}(t_1)\big) \ + \\ & \quad \quad \quad F_{l}\big(\mybar{\bm \alpha}(t_2),\ \mybar{\textbf{s}}(t_2)\big) - F_{l}\big(\mybar{\bm \alpha}(t_1),\ \mybar{\textbf{s}}(t_1)\big) \ + \\ & \quad \quad \quad \quad n^e_{l}(t_2) - n^e_{l}(t_1)\biggr] \\
		& = - \left[\Delta A_{l}\big(\Delta \mybar{\bm \alpha}\big) + \Delta F_{l}\big(\Delta \mybar{\textbf{s}}\big) + \Delta n^e_{l}\right]
	\end{aligned}
\end{equation}
where $\Delta A_{l}, \Delta F_{l}$ are respectively changes in shadowing and scattering losses due to the change in the DOI profile over $\Delta t$ and

\begin{equation}
	\label{Eq_DOIchange}
		\begin{aligned}
	 \Delta \mybar{\bm \alpha} &= \mybar{\bm \alpha}(t_2)- \mybar{\bm \alpha}(t_1)\\
	 \Delta \mybar{\textbf{s}} &= \mybar{\textbf{s}}(t_2)- \mybar{\textbf{s}}(t_1)
	 	\end{aligned}
\end{equation}

\subsection{Inverse Problem Formulation}
The imaging or inverse problem considered here can now be defined as estimating $\Delta \mybar{\bm \alpha}$ from measurements of the change in RSS across different links $\Delta P_{l}$.

If exact formulations are utilized, the relation between the change in RSS values and DOI spatial profile will be nonlinear \cite{Xudongchen}. Deriving these exact nonlinear models is difficult in practical conditions with an intricate background scattering environment. 

To avoid the difficulty of full wave models that are nonlinear, an alternative approach is to use approximate linear models, which have provided good experimental results \cite{dubeyTGRS, Patwari2010, Patwari2015}. Using the linear forms the indoor imaging inverse problem becomes
\begin{equation}
	\label{Eq_ConRTI111}
	\begin{aligned}
		\Delta P_{l} = - \big[\ \mybar{A}_{l}^T \Delta \mybar{\bm \alpha} + \mybar{F}_{l}^T \Delta \mybar{\textbf{s}} + \Delta n^e_{l}\big],
	\end{aligned}
\end{equation}
where, $\mybar{A}_{l} , \mybar{F}_{l} \in \mathcal{R}^{N\times 1}$ are respectively weight vectors derived from approximate linear shadowing and scattering models. $\Delta P_{l}$ is the measured change in RSS value for the $l_{m}^{th}$ link. The RSS data can then be measured for all $L=M(M-1)/2$ links to solve the system of linear equation (\ref{Eq_ConRTI111}) to estimate an averaged attenuation profile $\Delta \mybar{\bm \alpha}$ (the scattering profile $\Delta \mybar{\textbf{s}} \in \mathcal{R}^{N\times 1} $ can also be estimated but it has not been done for indoor RF imaging as explained later). 




The challenge is to find approximate linear models $\mybar{A}_{l} , \mybar{F}_{l} \in \mathcal{R}^{N\times 1}$ which can provide acceptable accuracy.

\section{Phaseless Indoor RF Imaging Techniques}

Using the framework (\ref{Eq_ConRTI111}) we define models for $\mybar{A}_{l} , \mybar{F}_{l}$ under RTI and the inverse scattering formulation, xPRA-LM. 


\subsection{RTI}
\label{Sec_RTI_form}
RTI assumes that the Wi-Fi signals propagate as straight rays along the LOS path between the transmitter and receiver. It further assumes that the signal only experiences shadowing/attenuation along the LOS path while more intricate wave scattering effects are ignored and treated as noise or distortion. The model for RTI, based on (\ref{Eq_ConRTI111}), can be expressed as
\begin{equation}
	\label{Eq_ConRTI3}
	\begin{aligned}
		\Delta P_{l} = - \mybar{A}_{l}^T \Delta \mybar{\bm \alpha} - {n}^{e,f}_{l},
	\end{aligned}
\end{equation}
where the noise becomes
\begin{equation}
	\label{Eq_ConRTIn}
	\begin{aligned}
		{n}^{e,f}_{l} = \mybar{F}_{l}^T \Delta \mybar{\textbf{s}} + \Delta {n}^e_{l}
	\end{aligned}
\end{equation}
In other words, while solving the RTI inverse problem, the $\Delta \mybar{\textbf{s}}$ term is ignored by treating it as noise so that we only solve for the attenuation term to image the DOI. 

From (\ref{Eq_ConRTI3}), the attenuation is defined such that the total attenuation experienced by the $l^{th}$ link is the weighted linear sum of attenuation caused by each grid and the weight of each grid is given by $\mybar{A}_{l}$. 

We can consider all $L=M(M-1)/2$ wireless links  (\ref{Eq_ConRTI3}) to be a system of linear equations,
\begin{equation} 
	\label{Eq_ConRTI_LS}
	\begin{aligned}
		\Delta \mybar{\textbf{P}} = - \mybar{\mybar{\textbf{A}}}_{\text{\tiny RTI}} \Delta \mybar{\bm \alpha}  + {\mybar{\textbf{n}}_\text{\tiny f,e}},
	\end{aligned}
\end{equation}
where, $\Delta \mybar{\textbf{P}} \in \mathbb{R}^{L\times 1}$ is the measured change in RSS vector for $L$ wireless links with elements $\Delta P_{l}, \ l = 1,2,...,L$. The image vector $\Delta \mybar{\bm \alpha} \in \mathcal{R}^{N\times 1}$ has entries $\left[\Delta \mybar{\bm \alpha}\right]_n, n=1,2,...,N$ representing the change in attenuation in each grid. The attenuation weight matrix $\mybar{\mybar{\textbf{A}}}_{\text{\tiny RTI}} \in \mathbb{R}^{L\times N}$ has elements $[\ \mybar{\mybar{\textbf{A}}}_{\text{\tiny RTI}}]_{{l, n}}$.
\begin{figure}[h]
	\centering
	\includegraphics[width=2in]{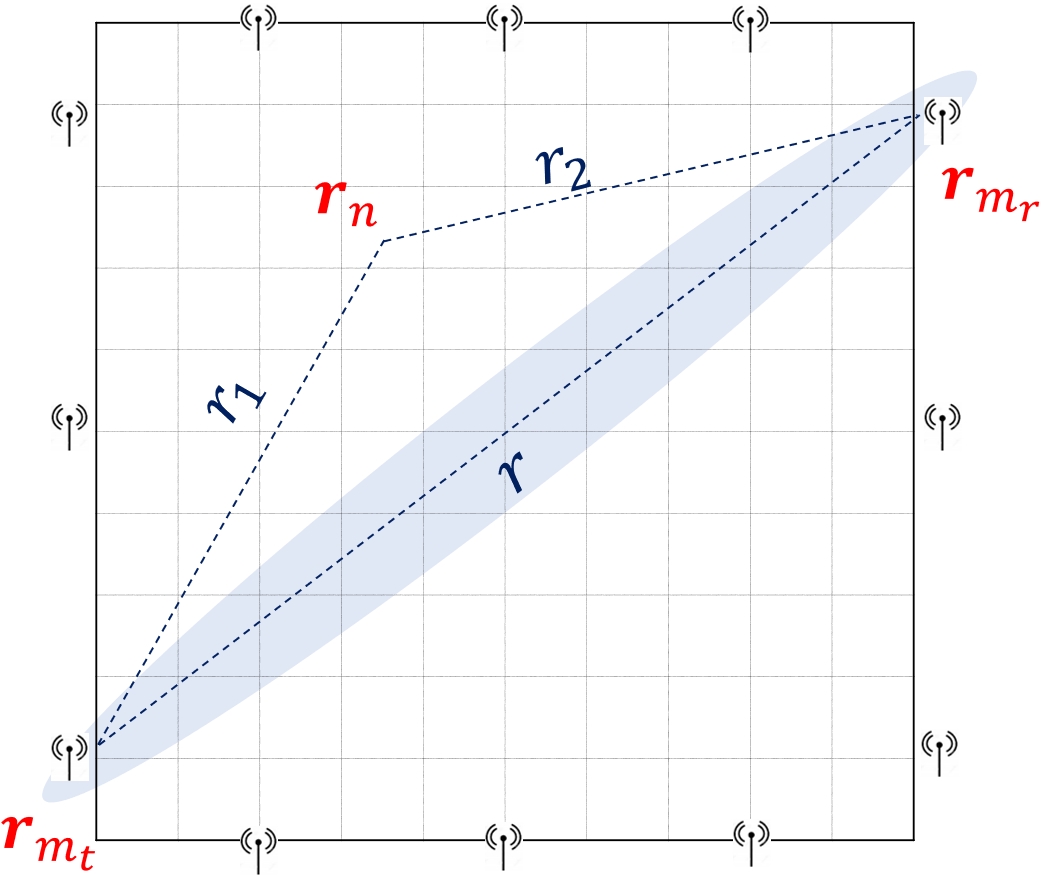}
	\caption{The DOI with wireless transceiver nodes at its boundary $ {\mathcal{B}}$. The transmitter $m_t$ and receiver $m_r$ are located at ${\bf r}_{m_t} \in {\mathcal{B}}$ and ${\bf r}_{m_r} \in {\mathcal{B}}$ respectively. The location of the $n^{th}$ grid in the DOI is denoted as ${\bf r}_n  \in {\mathcal{D}}$.}
	\label{RTInetwork3}
\end{figure}

To decide the structure of attenuation weight matrix $\mybar{\mybar{\textbf{A}}}_{\text{\tiny RTI}}$, RTI invokes a LOS assumption which implies that the signals travel along the LOS path and hence any attenuation is due to the grids which lie along the LOS path between transmitter and receiver. To enforce this, an ellipsoid method is used. As illustrated in Fig.\ref{RTInetwork3} an ellipse is considered with its foci at the transmitter and receiver. The grids which lie outside this ellipse are given zero weights. The grids which lie inside the ellipse are assigned a weight that is inversely proportional to the link distance $r = \left|\bm{r}_{m_t}-\bm{r}_{m_r}\right|$. Using this, the weights/entries of $\mybar{\mybar{\textbf{A}}}_{\text{\tiny RTI}}$ can be expressed as
\begin{equation}
	\label{Eq_weightRTI}
	[\ \mybar{\mybar{\textbf{A}}}_{\text{\tiny RTI}}]_{{l, n}} = 
	\begin{cases}
		\frac{1}{\sqrt{r}} &  \text{if} \ \ r_1+r_2 < r+\gamma\\
		0 & \text{otherwise}.
	\end{cases}       
\end{equation}
where, the distances $r, r_1$ and $r_2$ are given as (see Fig. \ref{RTInetwork3}),
\begin{equation}
	\label{Eq_weightRTI1}
	\begin{aligned}
		r = |\bm{r}_{m_t}-\bm{r}_{m_r}|; \quad
		r_1 = |\bm{r}_{m_t}-\bm{r}_{n}|; \quad
		r_2 = |\bm{r}_{n}-\bm{r}_{m_r}|
	\end{aligned}
\end{equation}
The tuning parameter $\gamma$ determines the width of the ellipse and is set such that the ellipse predominantly covers the grids along the LOS path \cite{Patwari2010}. In (\ref{Eq_weightRTI}), the inverse proportionality of weights to the link distance physically signifies that when the link is very long, the target object along the LOS path will have limited attenuation effect on the RSS, whereas when this link is short, then the objects along LOS can cause more variation in the RSS. This weighting strategy is based on the intuition and empirical studies and is the essence of the RTI formulation \cite{Patwari2010}.

\subsection{xPRA-LM}
In xPRA-LM, the extended form of the Rytov approximation is utilized to obtain an accurate linear model for the propagation loss and has been detailed in \cite{dubeyTGRS, DubeyTAP}. This subsection summarizes the development of xPRA-LM and complete details can be found in the original work \cite{dubeyTGRS}. One difference from \cite{dubeyTGRS} is that we also include the effect of temporal background subtraction here. 

Consider the $l^{th}$ wireless link with source at $\bm{r}_{m_t}$ and the receiver at $\bm{r}_{m_r}$ as shown in Fig. \ref{RTInetwork2}. The total electric field $E_{l}$ at the receiver (at $\bm{r}_{m_r}$) is
\begin{equation}
\label{Eq_Hzfree}
\begin{aligned}
E_{l} = E^i_{l}+ E^s_{l},
\end{aligned}
\end{equation}
where $E^i_{l}$ is the free-space incident field and $E^s_{l}$ is the scattered field in the presence of scatterers. Note that in the remainder of this section, the notation $E_{l}$ represents the field at the receiver $\bm{r}_{m_r}$ due to the transmitter at $\bm{r}_{m_t}$ whereas $E_{l}(\bm{r}_n)$ represents the field at the $n^{th}$ grid due to the transmitter at $\bm{r}_{m_t}$. 

The incident field and total field satisfies the free-space Helmholtz wave equation and inhomogeneous Helmholtz wave equations which can be respectively written as,
\begin{equation}
\label{Eq_Hzeqs}
\begin{aligned}
(\nabla^2 + k_0^2) E^i_{l}({\bf r}_n) = 0\ ;  \quad
(\nabla^2 + k_0^2 \nu^2({\bf r}_n)) E_{l}({\bf r}_n) = 0
\end{aligned}
\end{equation}
where $k_0=2\pi/\lambda_0$ is the free-space wavenumber. 

In xPRA-LM the Rytov transform is utilized \cite{dubeyTGRS, DubeyTAP} where the total field $E_l$ is written as
\begin{equation}
\label{Eq_RA}
E_{l}(\bm{r}) = E^i_{l}(\bm{r})  e^{\phi^s_{l}(\bm{r})},
\end{equation}
where the complex phase ${\phi^s_{l}}$ represents the complex wavefront deviations of the total field from the incident field. Note that (\ref{Eq_Hzfree}) and (\ref{Eq_RA}) are two different ways of expressing the total field. The phase term in the definition of (\ref{Eq_RA}) also includes attenuation and is a complex function and is known as the Rytov transform \cite{dubeyTGRS, wu2003wave}. There is no approximation introduced at this stage and (\ref{Eq_RA}) includes all the effects in the total field.

The complex phase in (\ref{Eq_RA}) is further partitioned as
\begin{equation}
\label{Eq_phi}
{\phi^s_{l}} = \phi^b_{l}(\bm{r}) + \phi^\Delta_{l}(\bm{r}),
\end{equation}
where $\phi^b_{l}$ is the wavefront resulting from the background stationary clutter such as walls, ceilings, floors and furniture and $\phi^\Delta_{l}(\bm{r})$ is the additional complex wavefront component due to the presence of the objects of interest inside the imaging region.

Substituting the definition of electric field (\ref{Eq_RA}) into the wave equations (\ref{Eq_Hzeqs}), we can obtain an exact integral description of wave scattering as shown in our recent work \cite{dubeyTGRS}. We refer to this integral as the Rytov Integral (RI) and it can be written as
\begin{equation}
\label{Eq_RIfinalcont}
\begin{aligned}
E_{l}   = E^i_{l} \cdot \exp\left( \frac{k_0^2}{E^i_{l}} \int_{\mathcal{D}}  g(\bm{r}_{m_r}, \bm{r}_n)  \chi_{\text{\tiny RI}}(\bm{r}_n)  E^i_{l}(\bm{r}_n) d\bm{r_n}^2 \right),
\end{aligned}
\end{equation} 
where $g(\cdot,\cdot)$ is the 2D Green’s function for the Helmholtz equation and $\chi_{\text{\tiny RI}}$ is RI's contrast given as,
\begin{equation}
\label{Eq_rytov2_chi}
\begin{aligned}
\chi_{\text{\tiny RI}} & (\bm{r}) = \nu(\bm{r})^2-1 + \frac{\nabla \phi_{l}^s(\bm{r}) \cdot \nabla \phi_{l}^s(\bm{r})}{k_0^2}.
\end{aligned} 
\end{equation}
This contrast function will be further simplified later. In discretized form, RI can be written as
\begin{equation}
\label{Eq_RIfinal2}
\begin{aligned}
E_{l}   = E^i_{l} \cdot \exp\left( \frac{k_0^2}{E^i_{l}} \sum_{\forall n}  g(\bm{r}_{m_r}, \bm{r}_n)  \chi_{\text{\tiny RI}}(\bm{r}_n)  E^i_{l}(\bm{r}_n) \Delta a \right),
\end{aligned}
\end{equation} 

One of the biggest advantages of the exponential form in (\ref{Eq_RIfinal2}) is that it can be transformed into a linear phaseless form. This can be performed by multiplying (\ref{Eq_RIfinal2}) by its conjugate and taking logarithms of both side to obtain,
\begin{equation}
\label{Eq_RytovInt}
\begin{aligned}
P_{l} & [\text{dB}]   = P_{l}^i [\text{dB}] \ +  \\ \ \  & C_0 \cdot \operatorname{Re}\bigg(\frac{k^2}{E^i_{l}}  \sum_{\forall n} g(\bm{r}_{m_r}, \bm{r}_n)  \chi_{\text{\tiny RI}} (\bm{r}_n) E^i_{l}(\bm{r}_n) \Delta a\bigg),
\end{aligned}
\end{equation}
where $C_0 = 20\log_{10}e$ is a constant. $P_{l} $ and $P^i_{l}$ are respectively the total received power (RSS) and free-space incident power in dB. 

Due to the linear relationship between contrast and measured RSS, we can straightforwardly include temporal background subtraction. Applying (\ref{Eq_NewConRTI2}) to (\ref{Eq_RytovInt}) we obtain
\begin{equation}
\label{Eq_RytovIntchange}
\begin{aligned}
\Delta P_{l} =C_0 \cdot \operatorname{Re}\bigg(\frac{k^2}{E^i_{l}}  \sum_{\forall n} g(\bm{r}_{m_r}, \bm{r}_n)   \Delta \chi_{\text{\tiny RI}} (\bm{r}_n) E^i_{l}(\bm{r}_n) \Delta a\bigg) ,
\end{aligned}
\end{equation}
where $\Delta  P_{l}$ is the change in RSS values (in dB) from instances $t_1$ to $t_2$ 
\begin{equation}
\label{Eq_xPRABsub}
\Delta P_{l} =  P_{l}({t_2} )- P_{l}({t_1})
=  20 \log_{10} \frac{|E_{l}({t_2}) |}{|E_{l}({t_1})|} 
\end{equation} 
and $\Delta \chi_{\text{\tiny RI}}$ is the change in the DOI RI contrast profile,
\begin{equation}
\Delta \chi_{\text{\tiny RI}}  = \chi_{\text{\tiny RI}}({t_2})-\chi_{\text{\tiny RI}} ({t_1}).
\end{equation}
{The xPRA-LM model in (\ref{Eq_RytovIntchange}) linearly relates the change in RSS values to the change in contrast profile and hence, can be used for imaging and tracking changes in a given indoor region.}

{It is important to understand the effect of the temporal background subtraction process on the contrast term (specifically on the complex phase term ${\nabla \phi_{l}^s(\bm{r}) \cdot \nabla \phi_{l}^s(\bm{r})}$). Let the background refractive index profile at time instant $t_1$ be $\nu^b$ and the associated complex phase of the total field be $\phi^b_{l}$. Then at time instant $t_2$, certain changes to the profile occurs and the refractive index profile becomes $\nu^s = \nu^b + \Delta \nu$, where, $\Delta \nu$ is the change in the profile. Let the associated complex phase of the total field be $\phi^s_{l} = \phi^b_{l} + \phi^\Delta_{l}$ (as explained in (\ref{Eq_phi})). Using (\ref{Eq_rytov2_chi}), we can now write $\Delta \chi_{\text{\tiny RI}}$ as}
\begin{equation}
\label{Eq_delta_phi}
\begin{aligned}
{\Delta \chi_{\text{\tiny RI}} } & = {(\nu^s)^2 - (\nu^b)^2 + \frac{1}{k^2_0} \left[\left(\nabla \phi_{l}^s \cdot \nabla \phi_{l}^s\right) - \left(\nabla \phi_{l}^b \cdot \nabla \phi_{l}^b\right)  \right]}\\
& = {(\nu^s)^2 - (\nu^b)^2 + \frac{1}{k^2_0} \left[{\nabla \phi_{l}^\Delta \cdot \nabla \phi_{l}^\Delta} + 2{\nabla \phi_{l}^b \cdot \nabla \phi_{l}^\Delta} \right].}
\end{aligned}
\end{equation}
{From (\ref{Eq_delta_phi}), we find that the complex phase term ${\nabla \phi_{l}^b \cdot \nabla \phi_{l}^b}$ associated with the scattering from background clutter is canceled. The two remaining terms are $2{\nabla \phi_{l}^b \cdot \nabla \phi_{l}^\Delta}$ (associated with the scattering between background clutter and objects of interests) and ${\nabla \phi_{l}^\Delta \cdot \nabla \phi_{l}^\Delta}$ (associated with the scattering due to the objects of interest). Here we can assume that the cross term $2{\nabla \phi_{l}^b(\bm{r}) \cdot \nabla \phi_{l}^\Delta(\bm{r})}$ is very small and can be neglected} based on the justification for background subtraction. That is the objects causing the background scattering in the indoor environment are usually spatially distinct from the scattering caused by the objects that are changing or moving and therefore the term  $2{\nabla \phi_{l}^b(\bm{r}) \cdot \nabla \phi_{l}^\Delta(\bm{r})}$ will be comparatively small to ${\nabla \phi_{l}^\Delta(\bm{r}) \cdot \nabla \phi_{l}^\Delta(\bm{r})}$. 


\subsubsection{Contrast Function Approximation}
{While the formulation for contrast, (\ref{Eq_delta_phi}), has been simplified using background subtraction it remains challenging to reconstruct it. In particular to utilize the contrast in (\ref{Eq_delta_phi}) for imaging, we need to simplify it to remove complex phase terms.} The term ${\nabla \phi_{l}^\Delta (\bm{r}) \cdot \nabla \phi_{l}^\Delta (\bm{r})}$ is unknown and is itself a function of the unknown change in refractive index profile of the DOI. To the best of our knowledge, the term ${\nabla \phi_{l}^\Delta (\bm{r}) \cdot \nabla \phi_{l}^\Delta (\bm{r})}$ has not been estimated for strong scattering conditions and hence the highly nonlinear inverse problem in (\ref{Eq_RIfinal2}) has not been solved. The straightforward approach to deal with this unknown gradient term is to neglect it, which gives well known Rytov approximation (RA). However, neglecting this term severely restrict the validity range of RA, making it futile for practical applications. In our recent work, instead of ignoring ${\nabla \phi_{l}^\Delta(\bm{r}) \cdot \nabla \phi_{l}^\Delta (\bm{r})}$, we approximate it using a high frequency approximation in lossy media (see \cite{dubeyTGRS} for details). This approach provides us with the final expression for a new approximate contrast $\chi_{{\nu}} \approx \chi_{\text{\tiny RI}}$ which is linearly proportional to refractive index (see \cite{dubeyTGRS} for the derivation),
\begin{equation}
\label{Eq_RIfinal2b}
\begin{aligned}
\chi_{{\nu}}  &=2 (\sqrt{\epsilon_R} -1)	+ j  \frac{\epsilon_I}{\sqrt{\epsilon_R}}\\ 
& = 2(\nu_R-1) +j 2\nu_I \quad \quad \text{for }\delta = \frac{\epsilon_I}{\epsilon_R} \ll 1\\
& = 2(\nu-1).
\end{aligned}
\end{equation} 
Note that the above approximate form of contrast is valid under the assumption that the scatterers exhibit low-loss. Nearly all objects in the indoor environment have a low-loss component and typical loss values are listed in \cite{komarov2005permittivity, Productnote, 4562803, dubeyTGRS}. In particular, the loss tangents of typical objects in an indoor environment, such as wood, concrete, and human body are of the order of 0.1 and this loss component should be included in formulation of indoor imaging and is another feature of xPRA-LM  \cite{dubeyTGRS}.

{Using this, we can now approximate the change in contrast $\Delta \chi_{\text{\tiny RI}}$ in (\ref{Eq_xPRABsub}b) with the change in contrast $\Delta \chi_{{\nu}}$ as,}
\begin{equation}
\label{Eq_delta_chi_nu}
\begin{aligned}
\Delta \chi_{{\nu}} &= 2 \Delta \nu  \quad && \left(\approx \Delta \chi_{\text{\tiny RI}}\right)  \\
& = 2 (\Delta \nu_R + j \Delta \nu_I).
\end{aligned}
\end{equation} 
Replacing the change in contrast $\Delta \chi_{\text{\tiny RI}} $ in (\ref{Eq_RytovIntchange}) with this new approximate change in contrast expression, the inverse problem is simpler as there is no need to estimate the intricate ${\nabla \phi_{l}^\Delta \cdot \nabla \phi_{l}^\Delta }$ term. 

The xPRA-LM model in (\ref{Eq_RytovIntchange}) is for a single wireless link. We can stack it for all $L = M(M-1)/2$ measurement links to obtain a linear system of equations,
\begin{equation}
\label{Eq_discrete1}
{\Delta \mybar{\textbf{P}} = \operatorname{Re}\big(\mybar{\mybar{\textbf{G}}} \  {\Delta \mybar{ \bm \chi}} \big)},
\end{equation}
where the measurement vector $\Delta \mybar{\textbf{P}} \in \mathbb{R}^{L\times 1}$ has elements $\Delta P_{l}, l=1,2,...,L$ [in dB]. The unknown contrast vector ${ \Delta \mybar{\bm \chi}}\in \mathbb{C}^{N\times 1}$ contains elements ${ \Delta {\chi_\nu}}(\bm{r}_n)$ for all $N$ grids inside DOI ($n = 1, 2,...,N$). The xPRA-LM model matrix ${\mybar{{\mybar{\textbf{G}}}}} \in \mathbb{C}^{L\times N}$ contains entries $\left[{\mybar{{\mybar{\textbf{G}}}}}\right]_{l,n}$ given as
\begin{equation}
\label{Eq_discrete2-x}
\left[\mybar{\mybar{\textbf{G}}}\right]_{l,n} =  \frac{C_0 k_0^2}{E^i_{l}}  g({\bf r}_{{m_r}}, \bm{r}_n)  E^i_{l}(\bm{r}_n) \Delta a
\end{equation}


Using above analysis, the xPRA-LM counterpart to RTI (\ref{Eq_ConRTI_LS}) can be written as
\begin{equation}
		\label{Eq_discrete3}
	\begin{aligned}
		\Delta \mybar{\textbf{P}} = -{\underbrace{\mybar{\mybar{\textbf{A}}}_{\text{\tiny xRA}} \Delta \mybar{\bm \alpha}}_{\substack{\text{ \scriptsize Change in} \\ \text{attenuation loss}}}  + \underbrace{\mybar{\mybar{\textbf{F}}}_{\text{\tiny xRA}} \Delta \mybar{\textbf{s}}}_{\substack{\text{ \scriptsize Change in}\\ \text{scattering loss}}}} + \ \mybar{\textbf{n}} _e
	\end{aligned}
\end{equation}
where, 
\begin{equation}
\label{Eq_discrete31}
\begin{aligned}
	{\mybar{\mybar{\textbf{A}}}_{\text{\tiny xRA}}} &  {= \operatorname{Im}(\mybar{\mybar{\textbf{G}}})/k_0; \quad
	\mybar{\mybar{\textbf{F}}}_{\text{\tiny xRA}} = \operatorname{Re}(\mybar{\mybar{\textbf{G}}})/k_0}\\
	\end{aligned}
\end{equation}
The xPRA-LM model matrix ${\mybar{{\mybar{\textbf{G}}}}} \in \mathbb{C}^{L\times N}$ contains entries $\left[{\mybar{{\mybar{\textbf{G}}}}}\right]_{l,n}$ given as
\begin{equation}
\label{Eq_discrete2}
\left[\mybar{\mybar{\textbf{G}}}\right]_{l,n} =  \frac{C_0 k_0^2}{E^i_{l}}  g({\bf r}_{{m_r}}, \bm{r}_n)  E^i_{l}(\bm{r}_n) \Delta a
\end{equation}
where constant $C_0 = 20\log_{10}e$, $k_0 = 2\pi/\lambda_0$ is the free-space wavenumber, $g(\cdot)$ is the homogeneous Greens function, $E^i_{l}$ and $E^i_{l}(\bm{r}_n)$ are the free-space incident electric fields respectively at the receiver and at any given point inside the DOI \cite{dubeyTGRS}. $\Delta a$ is the area of an individual grid.

Solving the regularized form of the xPRA-LM inverse problem in (\ref{Eq_discrete3}) can estimate $\Delta \mybar{\bm \alpha}$ using measurements $\Delta \mybar{\textbf{P}}$.   It has been shown \cite{dubeyTGRS} that xPRA-LM can provide attenuation profile reconstruction even under extremely strong scattering conditions and is shown to work for extremely high permittivity values $|\epsilon_r|\le 80$.


\section{Reconciling and Enhancing RTI}
\label{Sec_Analysis}
RTI is written as (\ref{Eq_ConRTI_LS}) while xPRA-LM is written as (\ref{Eq_discrete3}). These two methods differ in two major ways and these are: 1) the attenuation weight matrix $\mybar{\mybar{\textbf{A}}}_{\text{\tiny xRA}}$ is different in both its dependence on points in the DOI and also its overall magnitude and 2) the absence of scattering weight matrix $\mybar{\mybar{\textbf{F}}}_{\text{\tiny xRA}}$.

To reconcile the differences between the attenuation weight matrix in RTI and xPRA-LM, we first simplify $\mybar{\mybar{\textbf{A}}}_{\text{\tiny xRA}}$ so that it can be easily compared with RTI attenuation weight matrix $\mybar{\mybar{\textbf{A}}}_{\text{\tiny RTI}}$ in (\ref{Eq_ConRTI_LS}). 

The 2D Greens function and incident electric field for TM polarized waves can be expressed using Hankel functions and their corresponding asymptotic forms as
\begin{equation}
	\label{Eq_discrete6}
	\begin{aligned}
		&g(\bm{r}_{m_r}, \bm{r}_n) = \frac{j}{4} H^{(1)}_0(k_0 r_2)  \sim {\frac{a_0}{\sqrt{r_2}}} e^{j\left(k_0 r_2 +\frac{\pi}{4} \right)} \\
		&E^i_{l}(\bm{r}_n) = \frac{j}{4} H^{(1)}_0(k_0 r_1) \sim {\frac{a_0}{\sqrt{r_1}}} e^{j\left(k_0 r_1 +\frac{\pi}{4} \right)} \\
		&E^i_{l} = \frac{j}{4} H^{(1)}_0(k_0 r) \sim {\frac{a_0}{\sqrt{r}}} e^{j\left(k_0 r +\frac{\pi}{4} \right)} 
	\end{aligned}
\end{equation}
where, $a_0={\frac{1}{\sqrt{8\pi k_0}}}$ and $r_1, r_2$ and $r_3$ are defined in (\ref{Eq_weightRTI1}) and in Fig. \ref{RTInetwork3}. Using these we can expand the attenuation model weights of xPRA-LM as
\begin{equation}
	\label{Eq_discrete6a}
	\begin{aligned}
		{\left[\mybar{\mybar{\textbf{G}}}\right]_{l,n}} & {= \frac{C_0 k_0^2 \Delta a}{4} \left[j \frac{H^{(1)}_0(k_0 r_2)\ H^{(1)}_0(k_0 r_1)}{H^{(1)}_0(k_0 r)} \right]}
	\end{aligned}
\end{equation}

{Using the asymptotic forms in (\ref{Eq_discrete6}) (reasonably accurate for distances greater than $\lambda_0/2$ in our configuration), we can simplify the xPRA-LM attenuation weights $\left[\mybar{\mybar{\textbf{A}}}_{\text{\tiny xRA}}\right]_{l,n}$ as}
\begin{equation}
	\label{Eq_discrete7}
	\begin{aligned}
		\left[\mybar{\mybar{\textbf{A}}}_{\text{\tiny xRA}}\right]_{l,n} &= b_0 \sqrt{\frac{r}{r_1 r_2}} \sin\left(k_0\left[r_1+r_2-r\right] + \frac{\pi}{4}\right)
	\end{aligned}
\end{equation}
where $b_0= a_0 C_0 k_0 \Delta a$. 

%

For LOS paths (where $r_1+r_2 < r+\gamma$ and $\gamma \ll \lambda_0$), the argument of the $\sin$ function in (\ref{Eq_discrete7}) is approximately constant with value $1/\sqrt{2}$.  Therefore we can simplify (\ref{Eq_discrete7}) as
\begin{equation}
	\label{Eq_discrete8}
	\begin{aligned}
		\left[\mybar{\mybar{\textbf{A}}}_{\text{\tiny xRA}}\right]_{l,n} &= b_0 \sqrt{\frac{r}{2r_1 r_2}},
	\end{aligned}
\end{equation}
and hence, the xPRA-LM attenuation weights can be rewritten as
\begin{equation}
	\label{Eq_discrete9}
	\left[\mybar{\mybar{\textbf{A}}}_{\text{\tiny xRA}}\right]_{l,n} = 
	\begin{cases}
		b_0 \sqrt{\frac{r}{2 r_1 r_2}} &  \text{if} \ \ r_1+r_2 < r+\gamma \\
		b_0 \sqrt{\frac{r}{r_1 r_2}} \sin\left(k_0\left[\Delta r\right] + \frac{\pi}{4}\right) & \text{otherwise} .
	\end{cases}       
\end{equation}
where $\Delta r=r_1+r_2-r$.
Comparing (\ref{Eq_discrete9}) with LOS weights in RTI (\ref{Eq_weightRTI}), we can see that xPRA-LM and RTI have differences in their attenuation weight distribution. However, there are also similarities. In the middle of the LOS link where $r_1 \approx r_2 \approx r/2$ we can simplify the first row of (\ref{Eq_discrete9}) as
\begin{equation}
		\left[\mybar{\mybar{\textbf{A}}}_{\text{\tiny xRA}}\right]_{l,n} =  c_0 \sqrt{\frac{1}{r}},
\end{equation}
where $c_0= \sqrt{2} b_0$. This has a similar form to the first row of (\ref{Eq_weightRTI}). Therefore, near the middle of the LOS link, the $r$ dependence of RTI weights in (\ref{Eq_weightRTI}) can be justified using inverse scattering theory (used in xPRA-LM). To the best of our knowledge, none of the RTI-related work has provided this theoretical justification for the RTI weight dependence \cite{Patwari2010, Patwari2014, Patwari2017}. {However even with this $r$ dependence, the magnitude of the weights differs significantly. In particular, the term $c_0$ can be expanded as, }
\begin{equation}
	\label{Eq_constantxPRA}
	\begin{aligned}
	c_0 = \sqrt{2} b_0 = \frac{\left(20\log_{10}e\right) \Delta a }{\sqrt{2 \lambda_0}}.
	\end{aligned}
\end{equation}
{For 2.4 GHz incident field ($\lambda_0 = 0.125$ m) with a grid size of $\lambda_0/4$ (used later in results section), the value of term $c_0= 0.017$ is significantly different from unity as used in RTI. This, along with the second term in (\ref{Eq_discrete9}) (see next paragraph), is also another reason why the value of $\Delta \mybar{\bm \alpha}$ is not accurately estimated in RTI. }

Away from the mid point of the LOS link, we can see that the weight function for xPRA-LM significantly increases as $r_1$ or $r_2$ tend to zero and differs from RTI. This intuitively makes sense as objects near the transmitter or receiver (along LOS path) will tend to block the LOS link more significantly than objects in the center of LOS link. Furthermore, unlike RTI, in xPRA-LM, the attenuation weights are also assigned to non-LOS (NLOS) paths and is another critical reason why the value of $\Delta \mybar{\bm \alpha}$ is not accurately estimated in RTI. 

\begin{figure}[!ht]
	\captionsetup[subfigure]{margin={0.0cm,0cm}}
	\centering
	\begin{subfigure}{0.22\textwidth}
		\includegraphics[width=\textwidth]{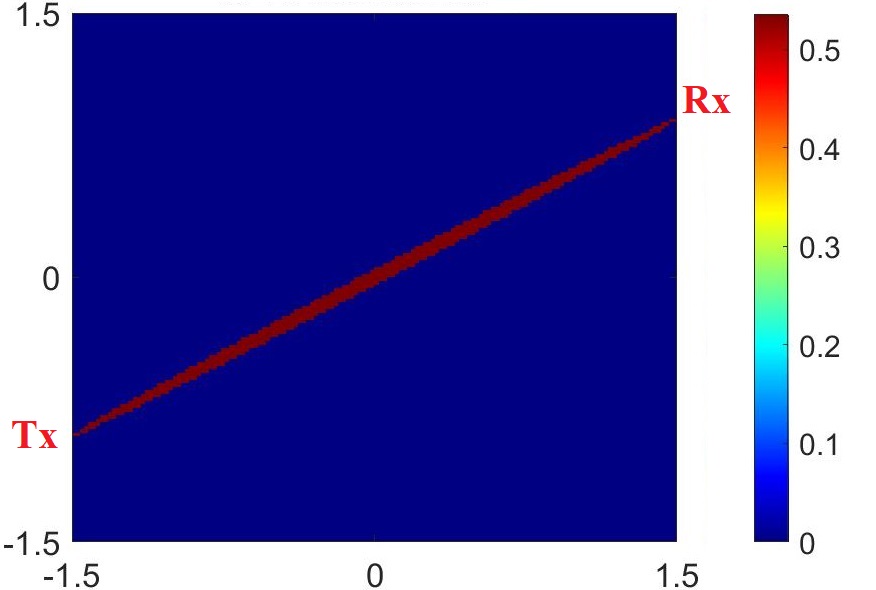}
		\subcaption{RTI}
	\end{subfigure}       
	\begin{subfigure}{0.22\textwidth}
		\includegraphics[width=\textwidth]{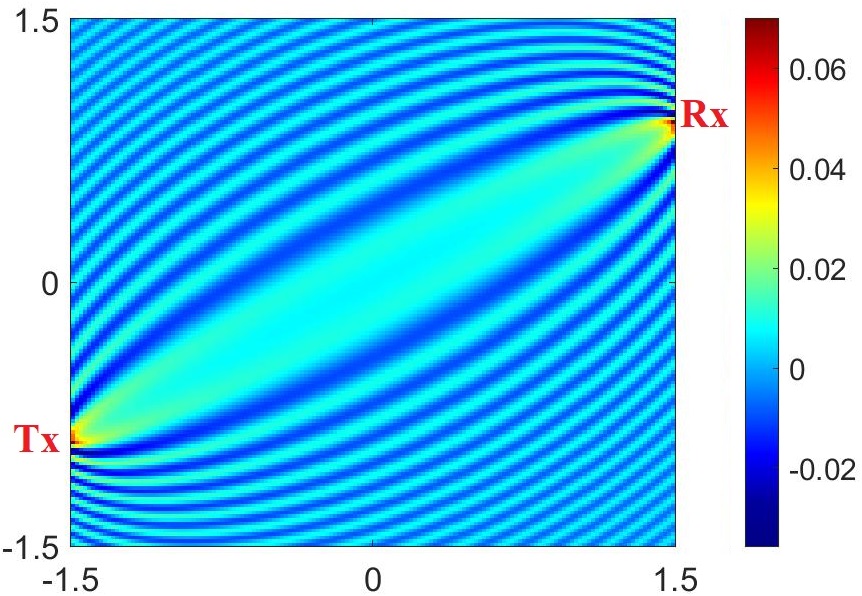}
		\subcaption{xPRA-LM}
	\end{subfigure}
	\caption{An illustration of the spatial distribution of RTI and xPRA-LM attenuation weights in a $3\times 3$ m$^2$ DOI (setup is similar to Fig. \ref{RTInetwork3}). The transmitter (Tx) and receiver (Rx) are located at coordinates $(-1.5, -0.9)$ and $(1.5, 0.9)$ respectively. The incident frequency used is 2.4 GHz.}
	\label{Fig_viskernel} 
\end{figure}

Fig. \ref{Fig_viskernel} illustrates the spatial distribution of the attenuation weights in a $3\times 3$ m$^2$ DOI (setup is similar to Fig. \ref{RTInetwork3}). The transmitter and receiver are located at coordinates $(-1.5, -0.9)$ and $(1.5, 0.9)$ respectively. The incident frequency used is 2.4 GHz. It can be seen that for RTI, the non-zero weights are only assigned to the LOS path, and the weights of all other NLOS grids are zero. The weights along the LOS path are inversely proportional to the distance between the transmitter and receiver. On the other hand, the spatial weight distribution in xPRA-LM is more intricate and takes the form of Fresnel zones of alternating zero and non-zero weight zones. Therefore, unlike RTI, xPRA-LM assigns weights to both LOS and NLOS paths. Furthermore from the scales of the sub-figures it can also be seen that the magnitude of the weights is different along the LOS paths of the two methods. 

{For the second difference between RTI and xPRA-LM,  consider the term $\left[\mybar{\mybar{\textbf{F}}}_{\text{\tiny xRA}}\right]_{l,n}$ which can be simplified to
	\begin{equation}
		\label{Eq_discrete7s}
		\begin{aligned}
			\left[\mybar{\mybar{\textbf{F}}}_{\text{\tiny xRA}}\right]_{l,n} &= b_0 \sqrt{\frac{r}{r_1 r_2}} \cos\left(k_0\left[r_1+r_2-r\right] + \frac{\pi}{4}\right).
		\end{aligned}
	\end{equation}
We can see that the only difference from $\left[\mybar{\mybar{\textbf{A}}}_{\text{\tiny xRA}}\right]_{l,n}$ is that the $\sin$ function has been changed to $\cos$. RTI does not have this term at all and lumps any scattering dependence $\Delta \mybar{\textbf{s}}$ into noise. In previous work \cite{dubeyTGRS}, it was shown that $\left[\mybar{\mybar{\textbf{F}}}_{\text{\tiny xRA}}\right]_{l,n}$  in xPRA-LM is accurate only under weak scattering conditions (when scatterer permittivity is small). For strong scattering conditions it has been shown that this scattering term is not accurate \cite{dubeyTGRS}. As discussed in \cite{dubeyTGRS} the reason for this is its dependence on the angle of the wavefronts inside the object and the term can in general by ignored.

\subsection{Proposed Enhancements to RTI}
Based on the observations made in the previous section, we propose straightforward enhancements to RTI to improve its performance. The most direct enhancement that can be performed to RTI is to replace the current RTI weight matrix (\ref{Eq_weightRTI}) with that from xPRA-LM (\ref{Eq_discrete7}). This provides an enhanced version of RTI as
\begin{equation} 
	\label{Eq_ConxRTI}
	\begin{aligned}
		{\Delta \mybar{\textbf{P}} = - \mybar{\mybar{\textbf{A}}}_{\text{\tiny xRA}} \Delta \mybar{\bm \alpha}  + {\mybar{\textbf{n}}_\text{\tiny f,e}},}
	\end{aligned}
\end{equation}
which we denote as xRTI in the remainder of this work. A further key simplification to xPRA-LM has been the removal of the term $\mybar{\mybar{\textbf{F}}}_{\text{\tiny xRA}}$. This simplification is motivated by the discussion at the end of the previous section (as also detailed in \cite{dubeyTGRS}) and it allows (\ref{Eq_ConxRTI}) to take a very similar form to the original RTI form (\ref{Eq_ConRTI_LS}). This form, (\ref{Eq_ConxRTI}), can allow the RTI community to straightforwardly try the new formulation.  
	

Overall, xRTI proposed in (\ref{Eq_ConxRTI}) can be interpreted in two ways. First, it can be seen as a modified version of RTI where we replace the empirical attenuation model in RTI by the inverse scattering based attenuation model from xPRA-LM (as explained above). The second way of interpreting (\ref{Eq_ConxRTI}) is that xRTI is an approximation to the original xPRA-LM model (\ref{Eq_discrete3}) where we ignore the scattering term $\mybar{\mybar{\textbf{F}}}_{\text{\tiny xRA}} \Delta \mybar{\textbf{s}}$. 

\section{Numerical and Experimental Results}
\label{Sec_Results}
This section provides simulation and experimental results for comparative analysis between RTI in (\ref{Eq_ConRTI_LS}) and xRTI in (\ref{Eq_ConxRTI}).

\subsection{Imaging Setup}
\label{Sec_setup}
Fig. \ref{problemsetup}(a) shows DOI setup for the simulation examples. The DOI size is $3 \times 3$ m$^2$ and 2.4 GHz Wi-Fi transceiver nodes are placed at the boundary of this DOI. These nodes can act as both sources and receivers. The setup utilizes a maximum of $M = 20$ identical transceiver nodes so that there are $L = M(M-1)/2 = 190$ unique links. Fewer nodes can be used (as shown later with $M=12$), but for the detailed comparative analysis of validity range, we use $M=20$ so that any errors are predominantly due to the model formulation and not because the problem is under-determined. 

Fig. \ref{problemsetup}(b) shows the experimental setup in a three-dimensional (3D) indoor environment where the goal is to image a 2D DOI cross section. To make the simulated and experimental results compatible, we make the geometry of DOI and sensor placement the same in both the simulation and experimental examples (more details on experimental setup are provided later). 
\begin{figure}[h]
	\centering
	\begin{subfigure}[t]{0.2\textwidth}
		\centering
		\includegraphics[width=\textwidth]{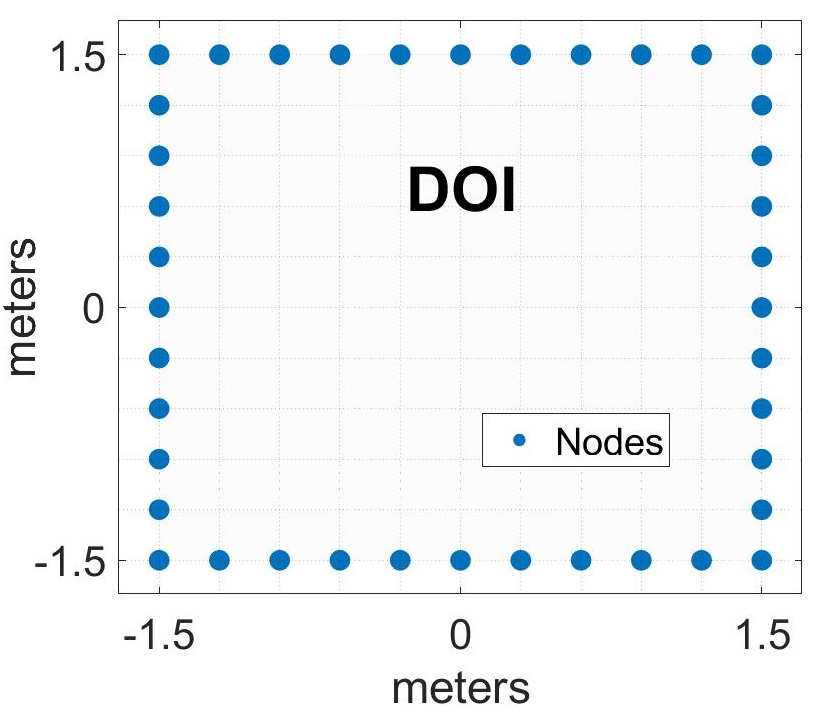}
		\subcaption{Simulation Setup}
	\end{subfigure}
	\begin{subfigure}[t]{0.27\textwidth}
		\centering
		\includegraphics[width=\textwidth]{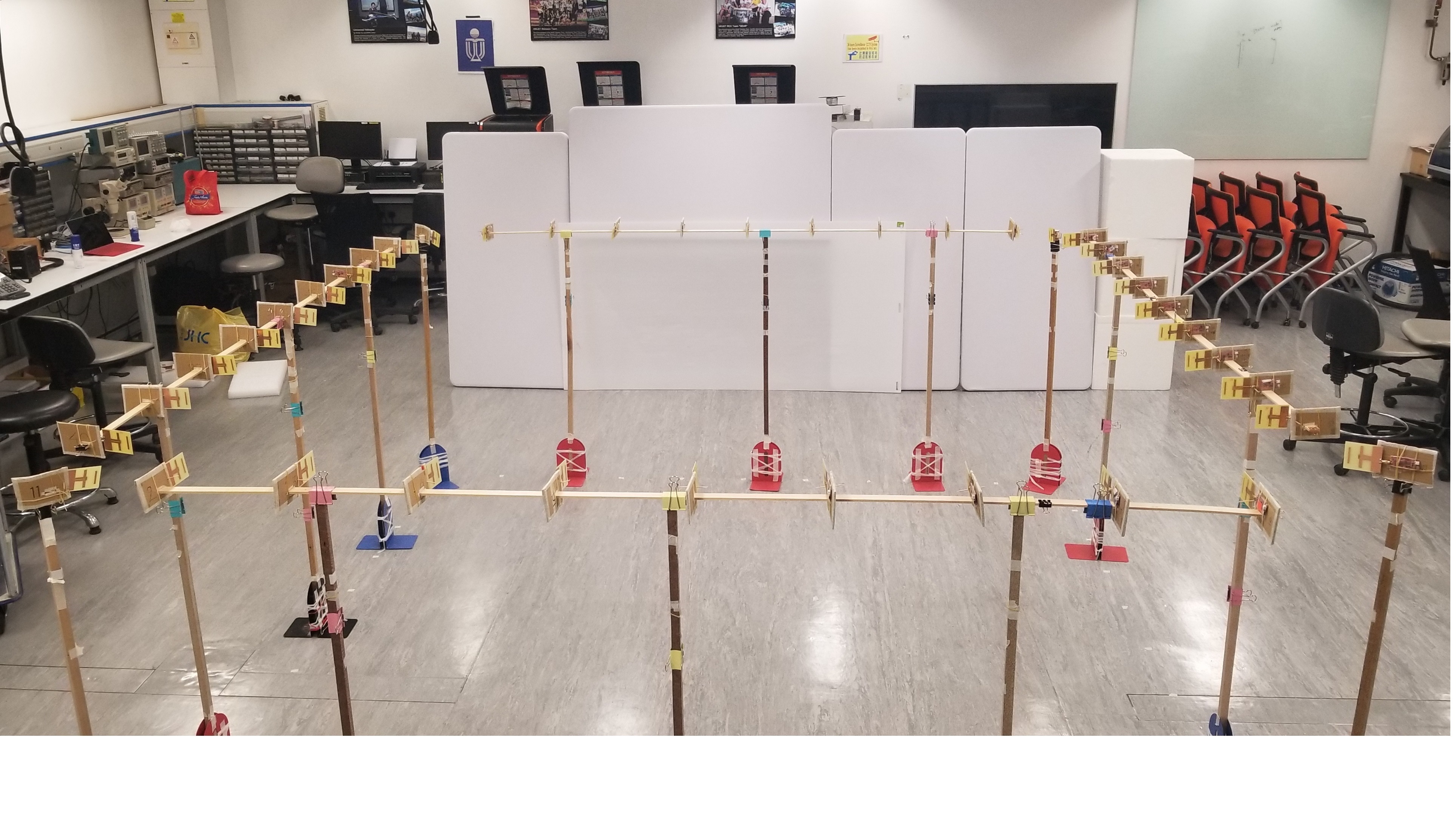}
		\subcaption{Experimental Setup}
	\end{subfigure}
	\caption{Imaging Setup. (a) Simulation setup where 2.4 GHz transceiver nodes (shown as blue dots) are placed at boundary of a $3\times 3$ m$^2$ DOI (b) Experimental setup in 3D environment including the 2D DOI with same configuration as simulation setup in (a).}
	\label{problemsetup}
\end{figure}

In the results, we focus on reconstructing two types of objects whose ground truth for the attenuation $\alpha$ is shown in Fig. \ref{Single_Scat_GT}. Note that we zoomed-in the $3\times 3$ m$^2$ DOI in Fig. \ref{Single_Scat_GT} to only show $1.5\times 1.5$ m$^2$ central area so that the scatterer can be seen clearly. 
Fig. \ref{Single_Scat_GT}a is a
circular lossy dielectric scatterer with diameter $40$ cm (or $3.2\times \lambda_0$) and is used for generating simulation results. The second scatterer profile has two scatterers with different permittivity (or attenuation values) values as shown in Fig. \ref{Single_Scat_GT}(b) and is used for simulation and experimental results.

In the simulation results for Fig. \ref{Single_Scat_GT}a, the value of $\epsilon_R$ ranges from 1.1 to 77 (corresponding value of $\alpha$ ranges from 5.2 to 44) and the exact values are specified in the simulation results section. For the object, we fix the loss-tangent $\delta=0.1$ so that the complex-valued relative permittivity of the test scatterers become $\epsilon_r = \epsilon_R+j\epsilon_I=\epsilon_R(1+0.1j)$. The corresponding attenuation parameter can be expressed as
\begin{equation}
	\label{Eq_alpha}
	\begin{aligned}
		\alpha = \frac{2\pi \epsilon_I}{\lambda_0 \sqrt{\epsilon_R}} = \frac{2\pi \delta \sqrt{\epsilon_R}}{\lambda_0}
	\end{aligned}
\end{equation}
For the objects in Fig. \ref{Single_Scat_GT}b, they consist of a square and cylinder with dimensions shown in the figure caption and its attenuation is fixed. For the simulations $\alpha=10$ and $\alpha=15.8$ while for the experiments (shown later in this section), the attenuation values are $\alpha=6.8$ and $\alpha=44$.

\begin{figure}[h]
	\centering
	\begin{subfigure}[t]{0.15\textwidth}
		\centering
		\includegraphics[width=\textwidth]{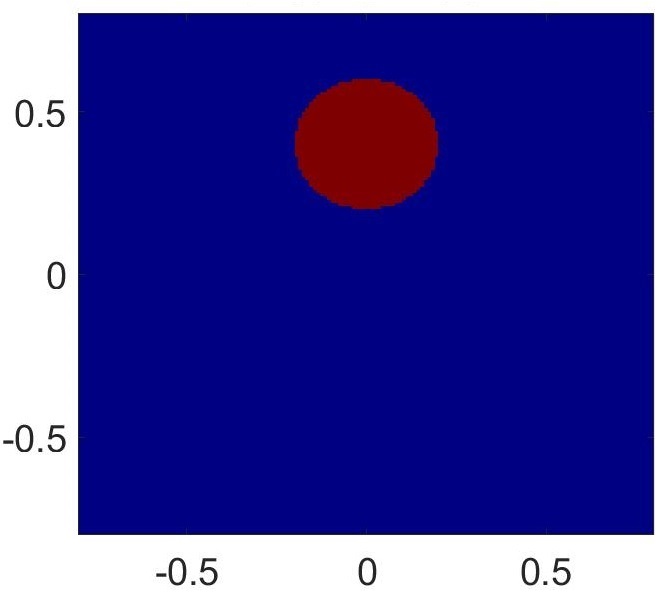}
		\caption{}
	\end{subfigure}
	\begin{subfigure}[t]{0.18\textwidth}
		\centering
		\includegraphics[width=\textwidth]{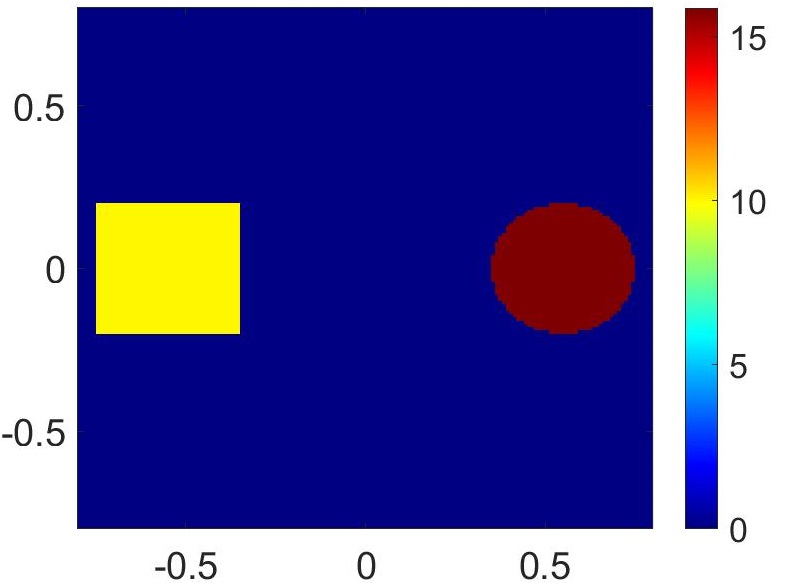}
		\caption{}
	\end{subfigure}
	\caption{Ground truth for objects used in the simulations (a) Circular lossy dielectric object with diameter of $40$ cm (or $3.2 \lambda_0$) and its attenuation parameter profile is provided in (\ref{Eq_alpha}) (in this figure it is 15.8).  (b) Object composed of a square ($\alpha=10$ and $\epsilon_r=4+j0.4$) and circular ($\alpha=15.8$ and $\epsilon_r=10+j1$)  object. The square and cylinder have edges and a diameter of $40$ cm respectively. Note that the original DOI is $3\times 3$ m$^2$ but DOI shown in this figure is magnified to clearly show the scatterers.}
			\label{Single_Scat_GT}
\end{figure}

\begin{figure}[h!]
	\captionsetup[subfigure]{justification=centering}
	\begin{subfigure}[b]{1\linewidth}
		\centering	
		\adjustbox{minipage=0.4\linewidth}{\subcaption*{\textbf{{\normalsize RTI}} }} \hspace{10mm}
		\adjustbox{minipage=0.4\linewidth}{\subcaption*{\textbf{{\normalsize xRTI}} }} \\
		\includegraphics[width=0.4\linewidth]{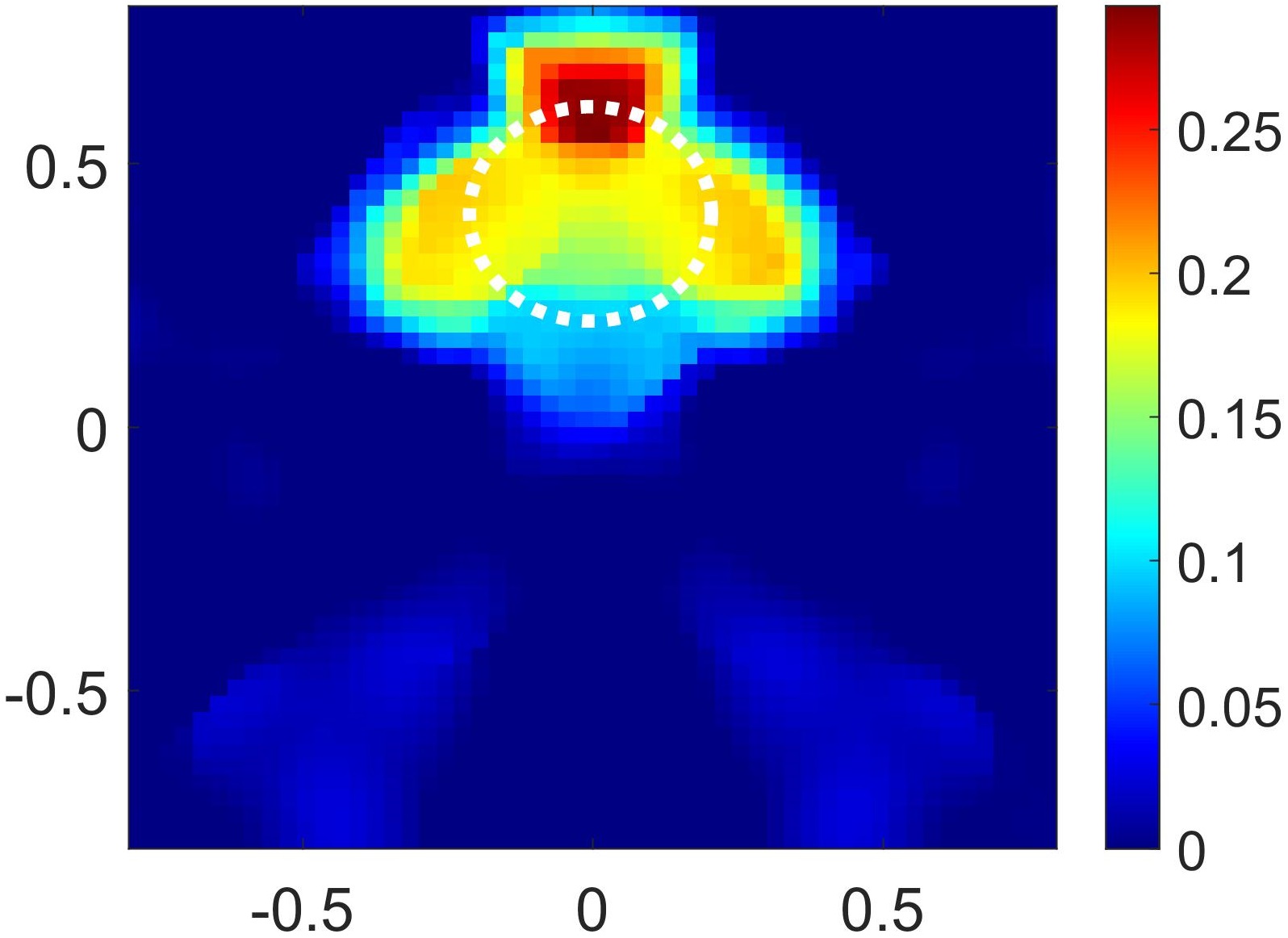} \hspace{10mm}
		\includegraphics[width=0.4\linewidth]{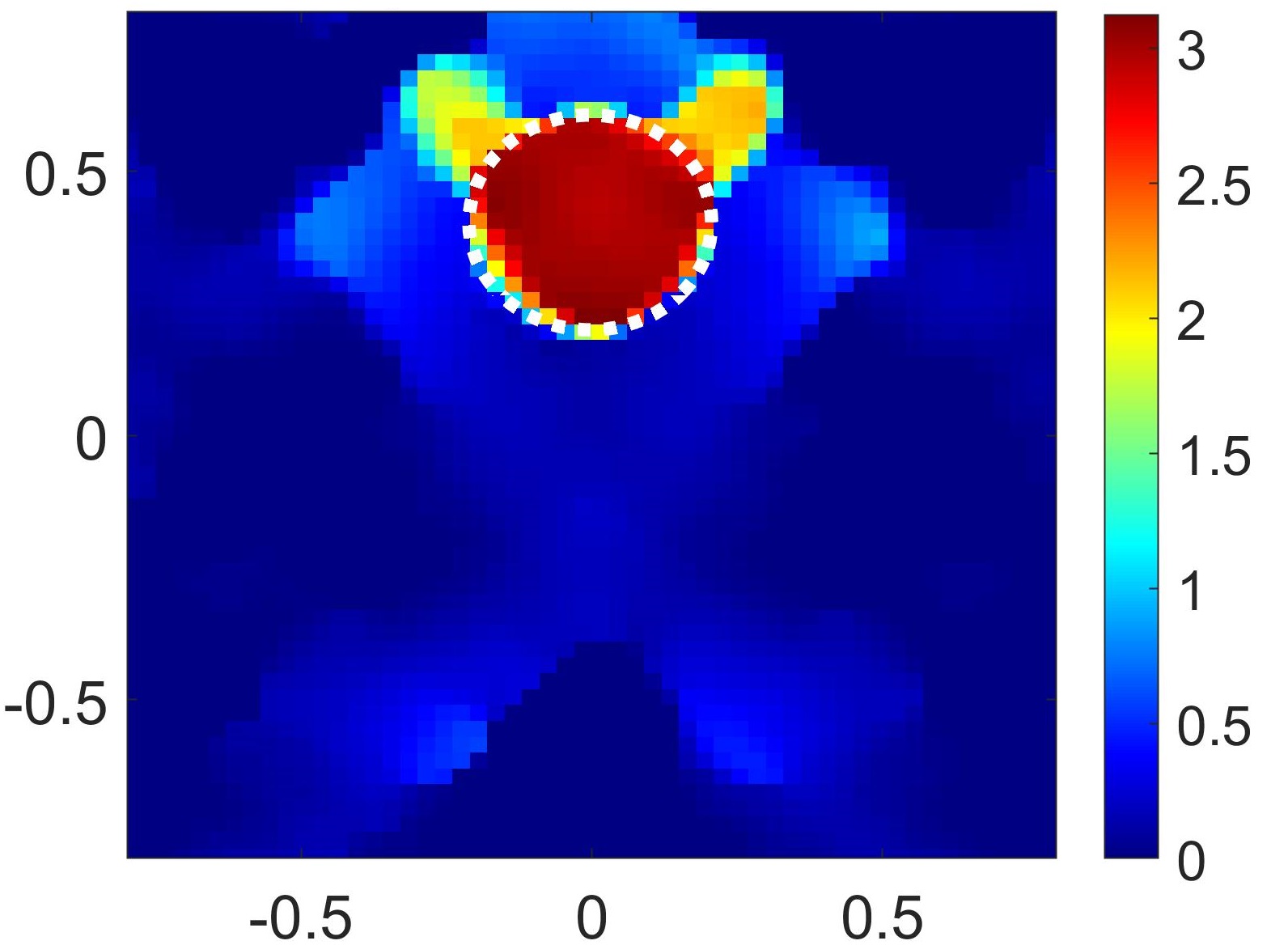}
		\caption{Reconstruction results when $\bm{ \alpha = 5.2}$ (or $\bm{ \epsilon_r=1.1+j0.11}$). The PSNR values for RTI and xRTI are 13.3 dB and 19.4 dB respectively. The corresponding values of $p$ (in regularization parameter $\Phi=2^p$) are 3 and 4.}
	\end{subfigure} 
	\vspace{0.2\baselineskip}
	\begin{subfigure}[b]{1\linewidth}
		\centering
		\includegraphics[width=0.4\linewidth]{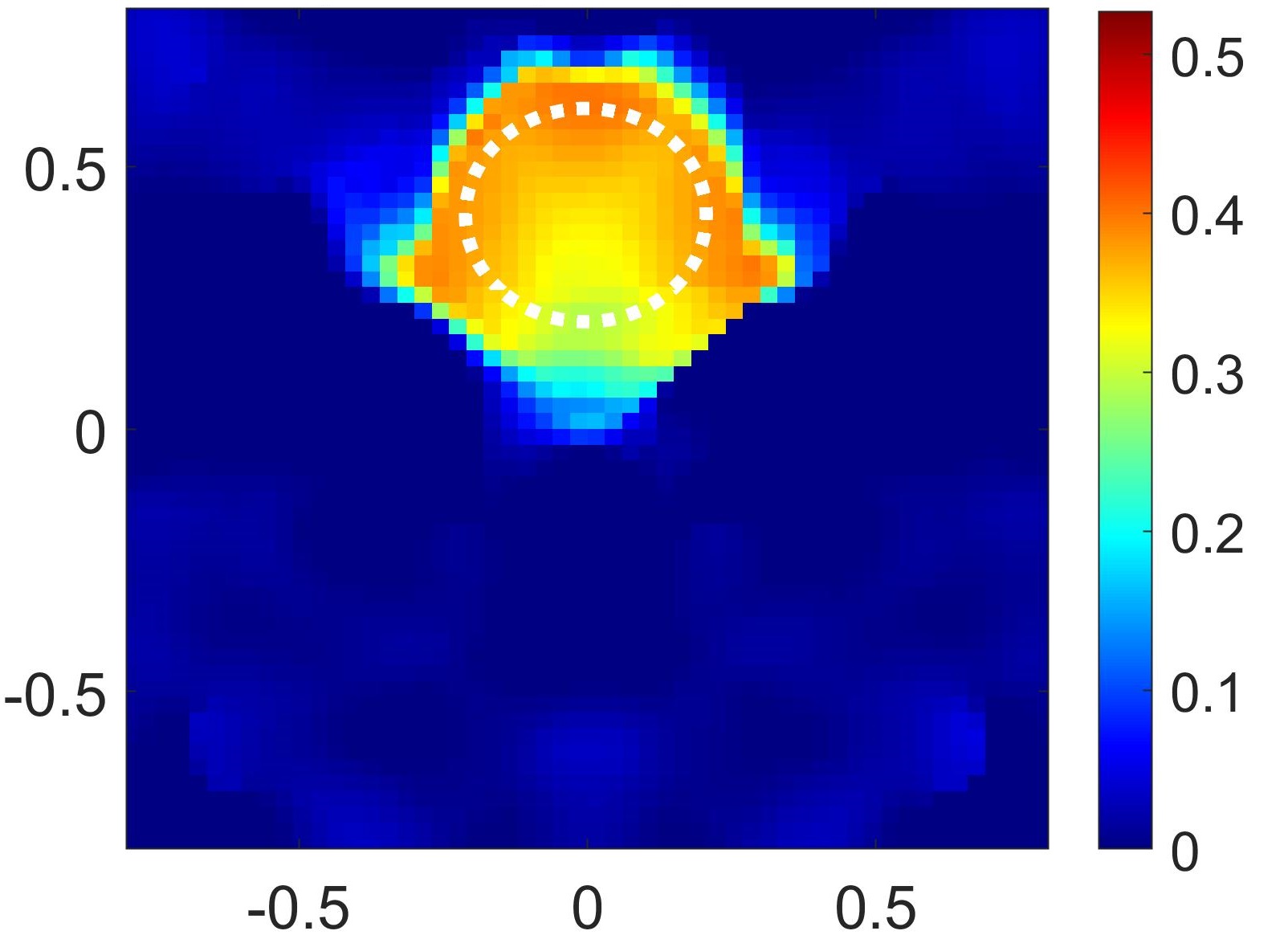} \hspace{10mm}
		\includegraphics[width=0.4\linewidth]{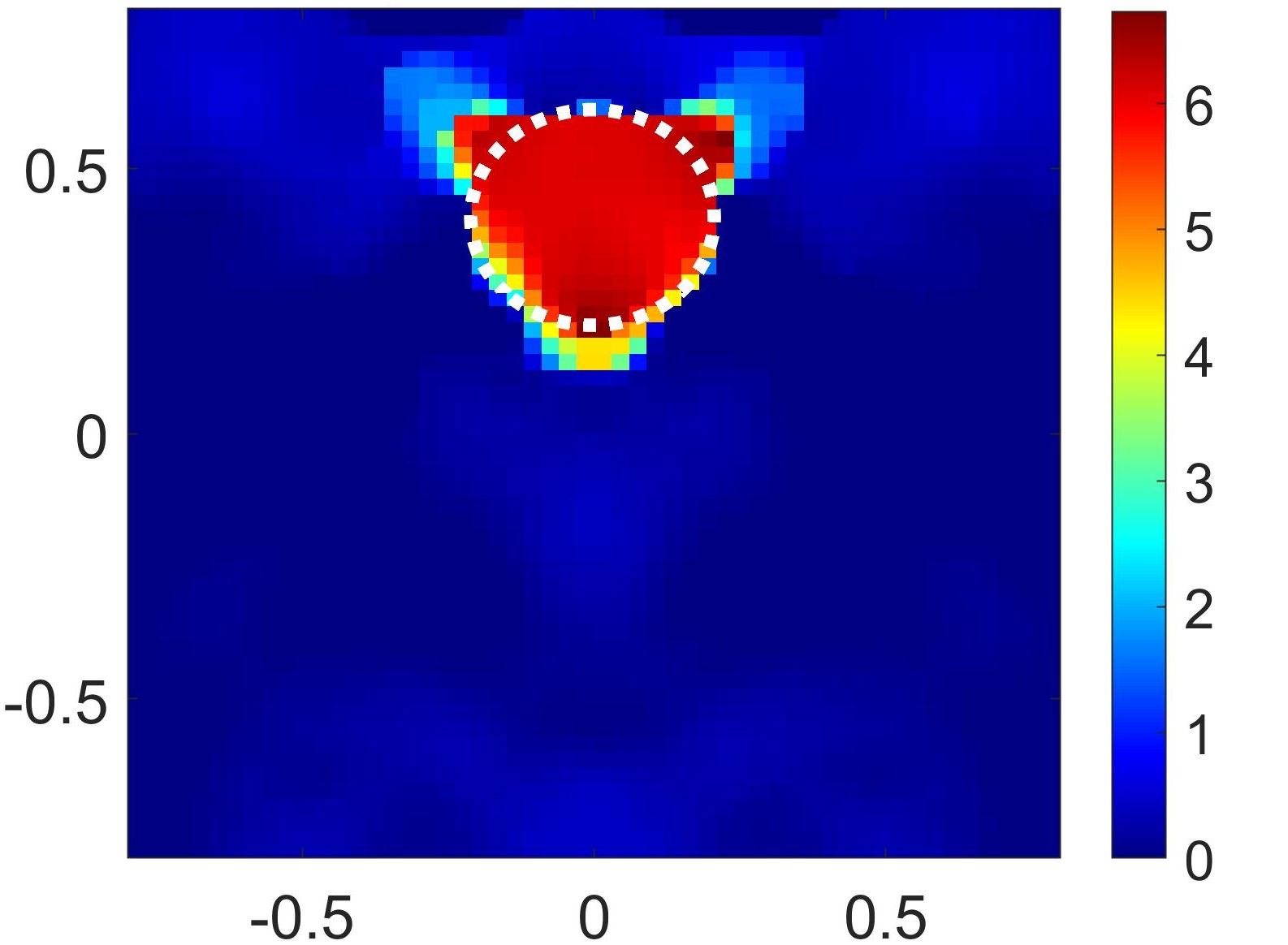}
		\caption{Reconstruction results when $\bm{\alpha = 7.1}$ (or $\bm{ \epsilon_r=2+j0.2}$). The PSNR values for RTI and xRTI are 13.2 dB and 19.6 dB respectively. The corresponding values of $p$ (in regularization parameter $\Phi=2^p$) are 3 and 4.}
	\end{subfigure} 
	\vspace{0.2\baselineskip}
	\begin{subfigure}[b]{1\linewidth}
		\centering
		\includegraphics[width=0.4\linewidth]{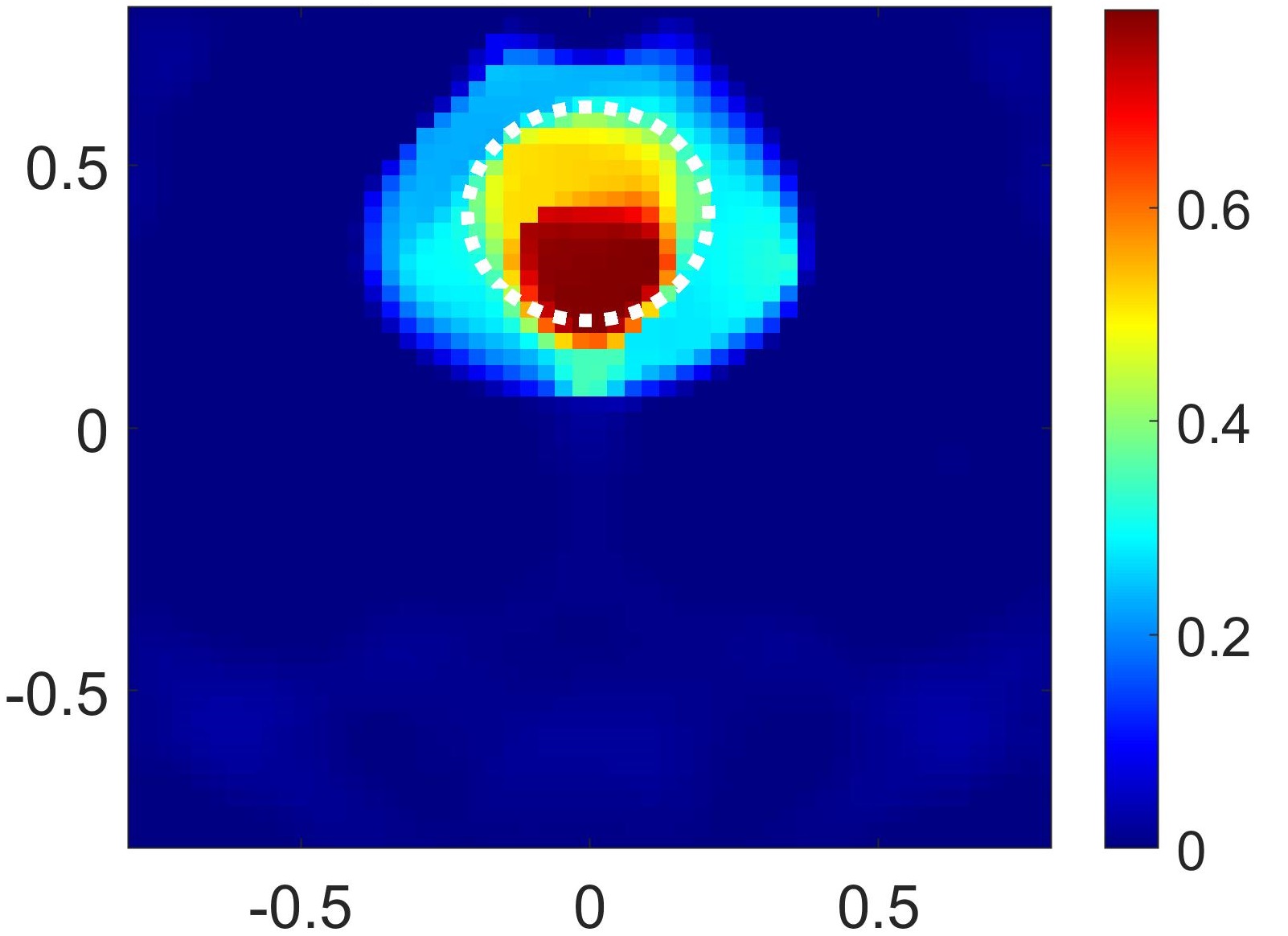} \hspace{10mm}
		\includegraphics[width=0.4\linewidth]{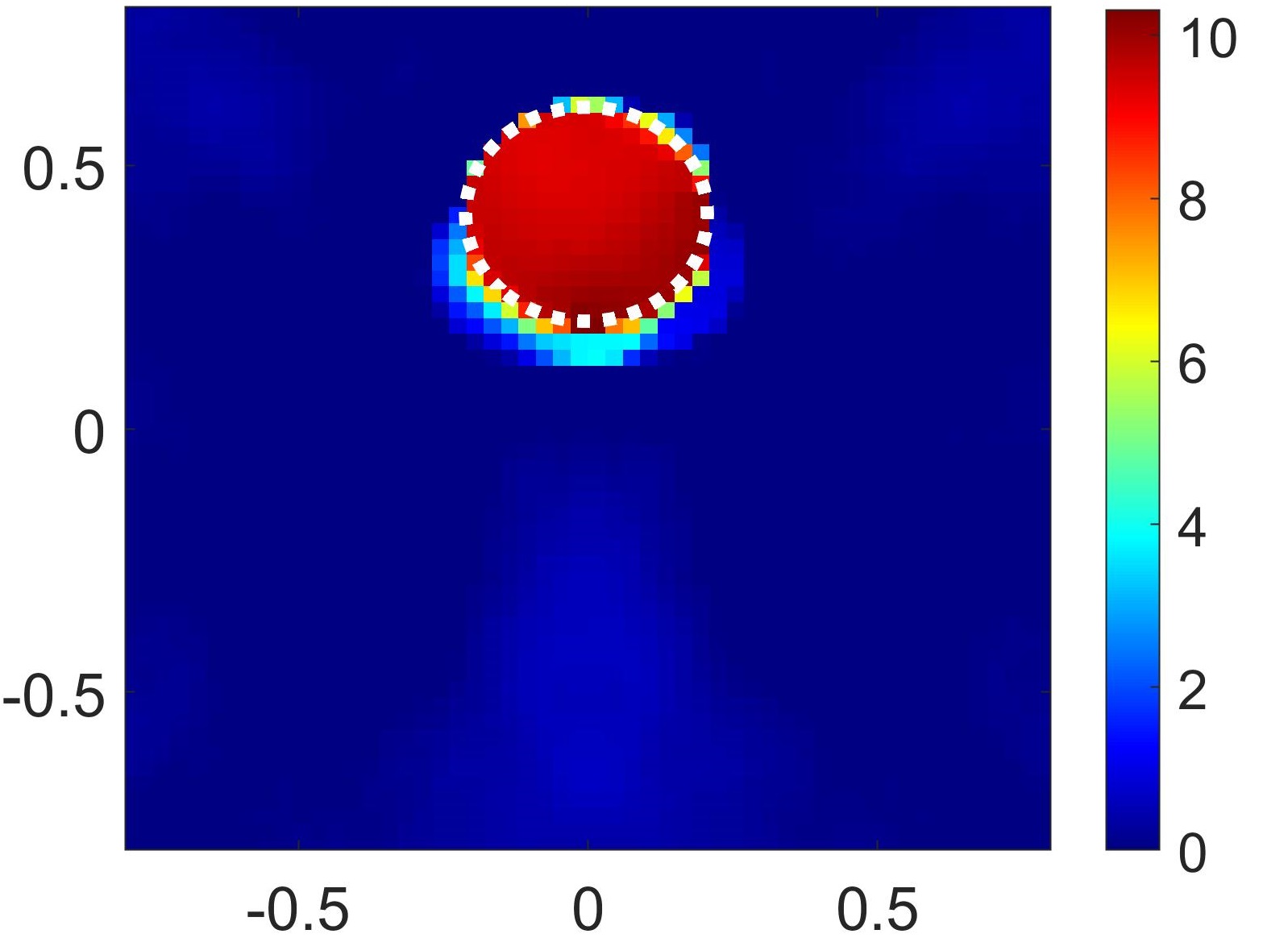}
		\caption{Reconstruction results when $\bm{\alpha = 10}$ (or $\bm{ \epsilon_r=4+j0.4}$). The PSNR values for RTI and xRTI are 13.8 dB and 21 dB respectively. The corresponding values of $p$ (in regularization parameter $\Phi=2^p$) are 3 and 4.}
	\end{subfigure} 
	\vspace{0.2\baselineskip}
	\begin{subfigure}[b]{1\linewidth}
		\centering
		\includegraphics[width=0.4\linewidth]{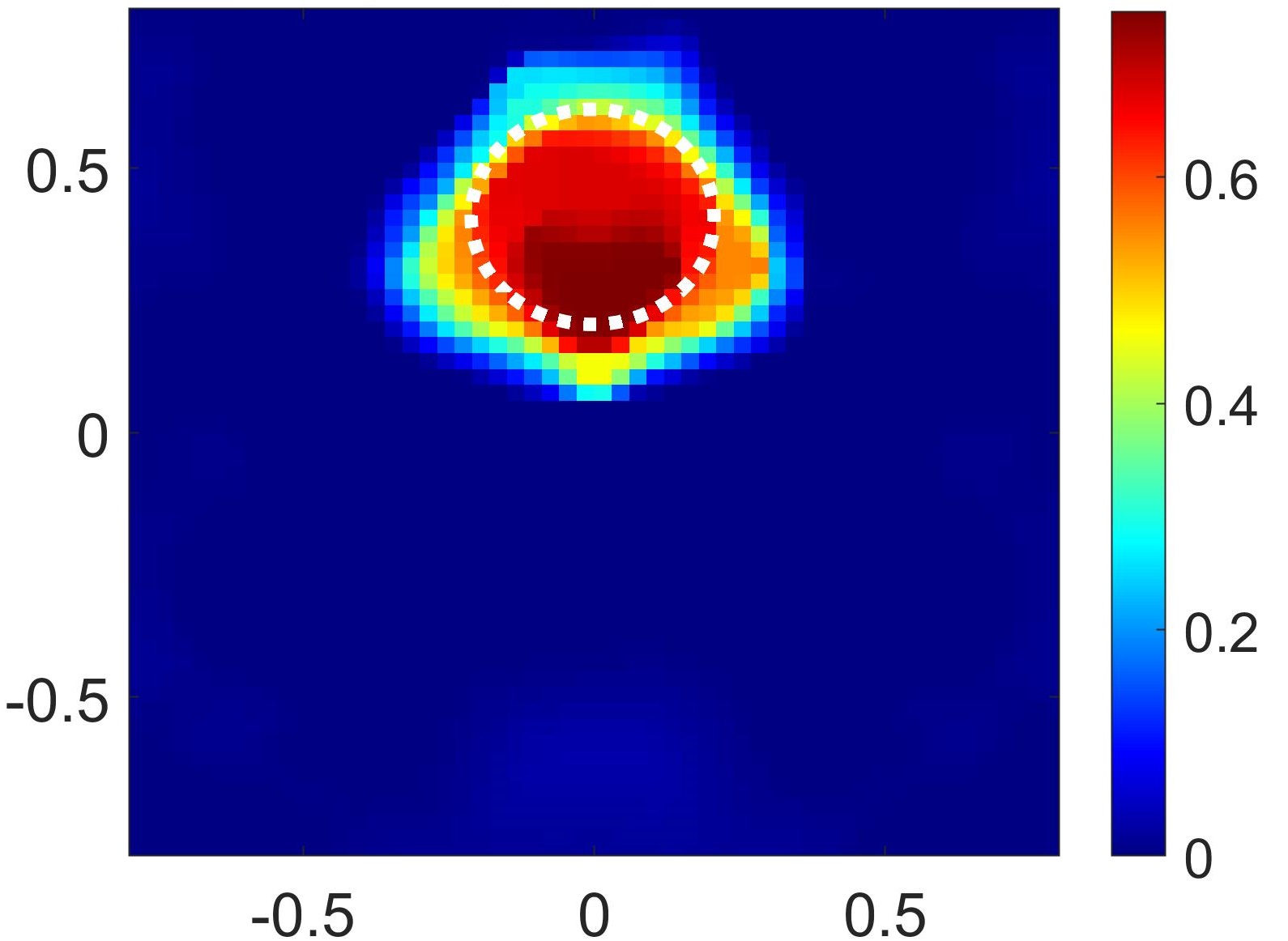} \hspace{10mm}
		\includegraphics[width=0.4\linewidth]{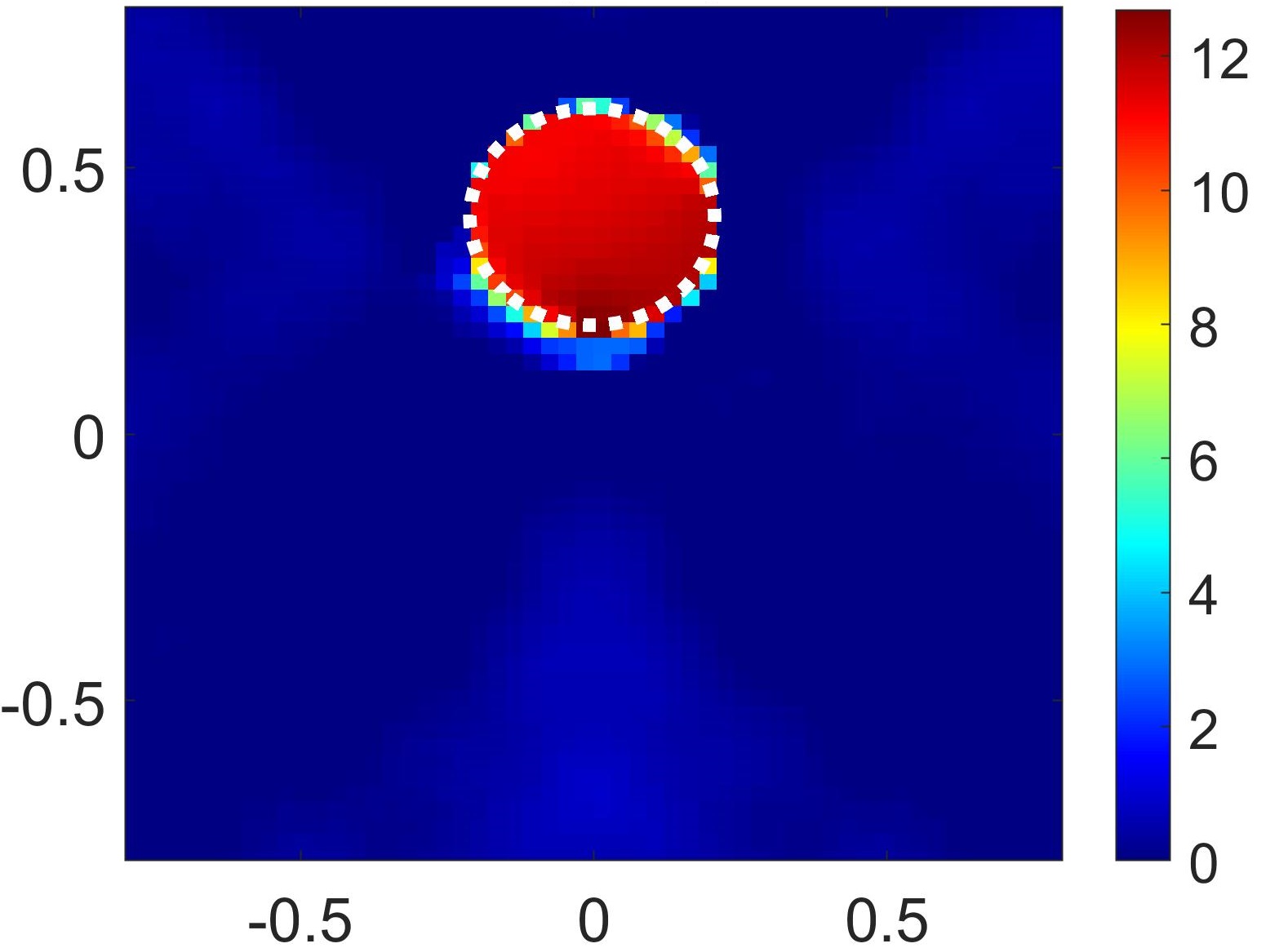}
		\caption{Reconstruction results when $\bm{\alpha = 15.8}$ (or $\bm{ \epsilon_r=10+j1}$). The PSNR values for RTI and xRTI are 13.7 dB and 22 dB respectively. The corresponding values of $p$ (in regularization parameter $\Phi=2^p$) are 3 and 6.}
	\end{subfigure} 
	\vspace{0.2\baselineskip}
	\begin{subfigure}[b]{1\linewidth}
		\centering
		\includegraphics[width=0.4\linewidth]{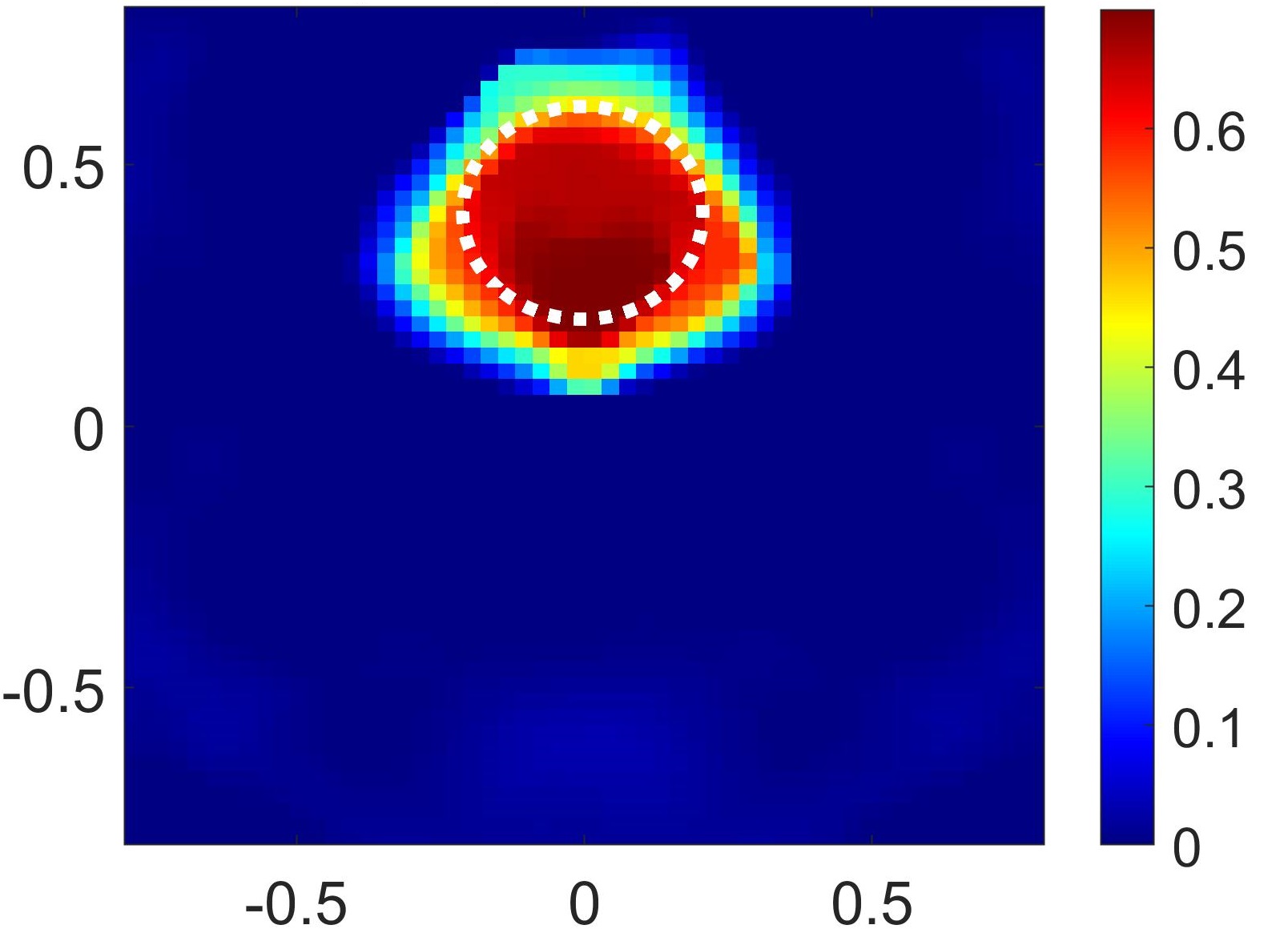} \hspace{10mm}
		\includegraphics[width=0.4\linewidth]{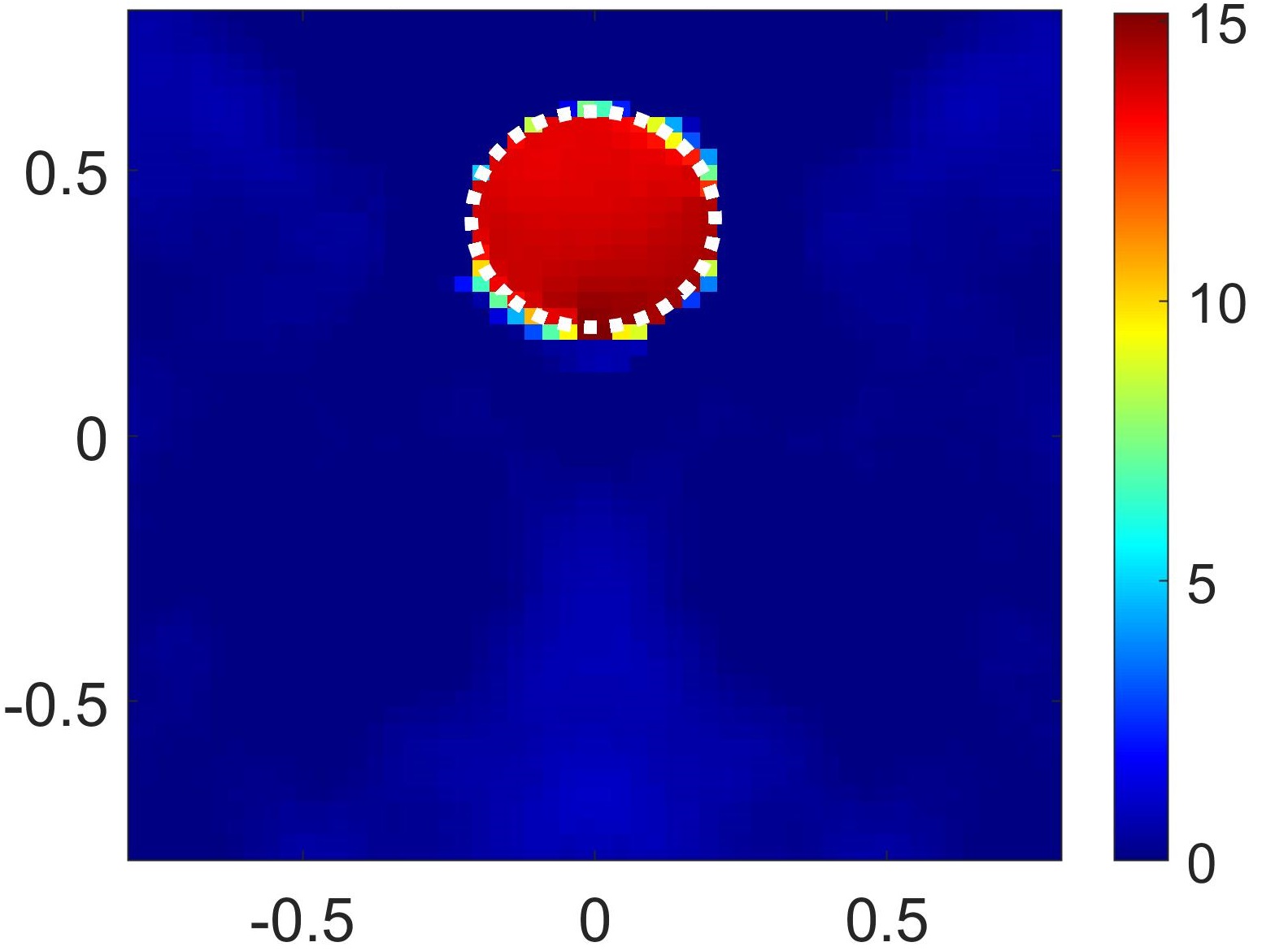}
		\caption{Reconstruction results when $\bm{\alpha = 44}$ (or $\bm{ \epsilon_r=77+j7}$). The PSNR values for RTI and xRTI are 13.2 dB and 16.2 dB respectively. The corresponding values of $p$ (in regularization parameter $\Phi=2^p$) are 3 and 6.}
	\end{subfigure}   		
	\caption{Using RTI and xRTI to reconstruct the ground truth attenuation profile shown in Fig. \ref{Single_Scat_GT}a for the varying values of attenuation parameter (or the relative permittivity). The color-scale shown in the results represent the attenuation parameter $\alpha= {2\pi \delta \sqrt{\epsilon_R}}/{\lambda_0}$. The x and y-axis are in meters.}
	\label{Single_scat} 
	\vspace{-0.1\baselineskip}
\end{figure}

For solving the inverse imaging problem, the $3 \times 3$ m$^2$ DOI in Fig. \ref{problemsetup}(a) is divided into $3 \times 3$ cm$^2$ discrete grids (grid size in terms of wavelength is approximately $\frac{\lambda_0}{4} \times \frac{\lambda_0}{4}$, where $\lambda_0=12.5$ cm for 2.4 GHz Wi-Fi). Hence we need to estimate $N = 100\times 100 = 10000$ unknowns using $190$ measurements which is a severely under-determined problem. 

To solve the ill-posed imaging problem using RTI (\ref{Eq_ConRTI_LS}) and xRTI (\ref{Eq_ConxRTI}), {we regularize the least square objective of the model with a smoothness prior which minimizes the successive difference in the unknown coefficients (in horizontal and vertical direction) to obtain a sparse and smooth reconstruction. This can be done using well-known total variation (TV) regularization \cite{li2010efficient, li2009user}. We use isotropic TV (with norm-2 prior term) and solve it with the alternating direction method of multipliers as implemented in TVAL3 package (see documentation and explanation in \cite{li2010efficient, depatla2015x, li2009user}). } {The regularization parameter in TVAL3 \cite{li2009user} is given by $\Phi = 2^p$ where $p$ is any real number. Note that the reconstructions shown in all the results are obtained by using the value of regularization parameter $\Phi = 2^p$ that achieves the highest PSNR. This value of regularization parameter can be different for RTI and xRTI. The value of $\Phi$ used for each of the reconstructions are provided in the captions of the corresponding reconstruction figures.}

It is important to note that using our straightforward classical prior instead of an advanced regularization (such as data-driven priors \cite{Unrolled, DeshmukhTAP}) helps us make sure that the reconstruction accuracy is related to the RTI and xRTI models and is not achieved using advanced regularization techniques that can overcomes the limitations of the models.  

To generate simulation RSS data $\Delta \bar{\textbf{P}}$, the exact forward problem is solved using the method of moments (MoM) technique (see \cite{chen2018computational, dubeyTGRS} for details). Since this technique is independent of RTI and xRTI, it avoids the possibility of an "inverse crime"  \cite{chen2018computational, dubeyTGRS} where the same model is used for date generation and also reconstruction. 

\subsection{Simulation Results}


Fig. \ref{Single_scat} provides reconstruction results for the test scatterer (in Fig. \ref{Single_Scat_GT}(a)) with different permittivity values. These are obtained using RTI and xRTI as shown in the first, second and third column respectively in Fig. \ref{Single_scat}. Fig. \ref{Single_scat}(a), (b), (c), (d) and (e) which provide attenuation profile reconstructions when the real part of relative permittivity is set to $\epsilon_R=1.1, 2, 4, 10$ and $77$ respectively and loss tangent is fixed at $\delta = 0.1$. This range of relative permittivity covers an extremely large range of scattering conditions and $\epsilon_R=10$, $77$ simulates extremely strong scattering conditions which is beyond any existing state-of-the-art nonlinear phaseless inverse scattering methods \cite{Xudongchen, dubeyTGRS, chen2018computational}. The reconstructions in Fig. \ref{Single_scat}(a)-(e) are provided in terms of the attenuation parameter (which is related to permittivity using (\ref{Eq_alpha})). 

\begin{figure}[!h]
	\centering
	\includegraphics[width=2.1in]{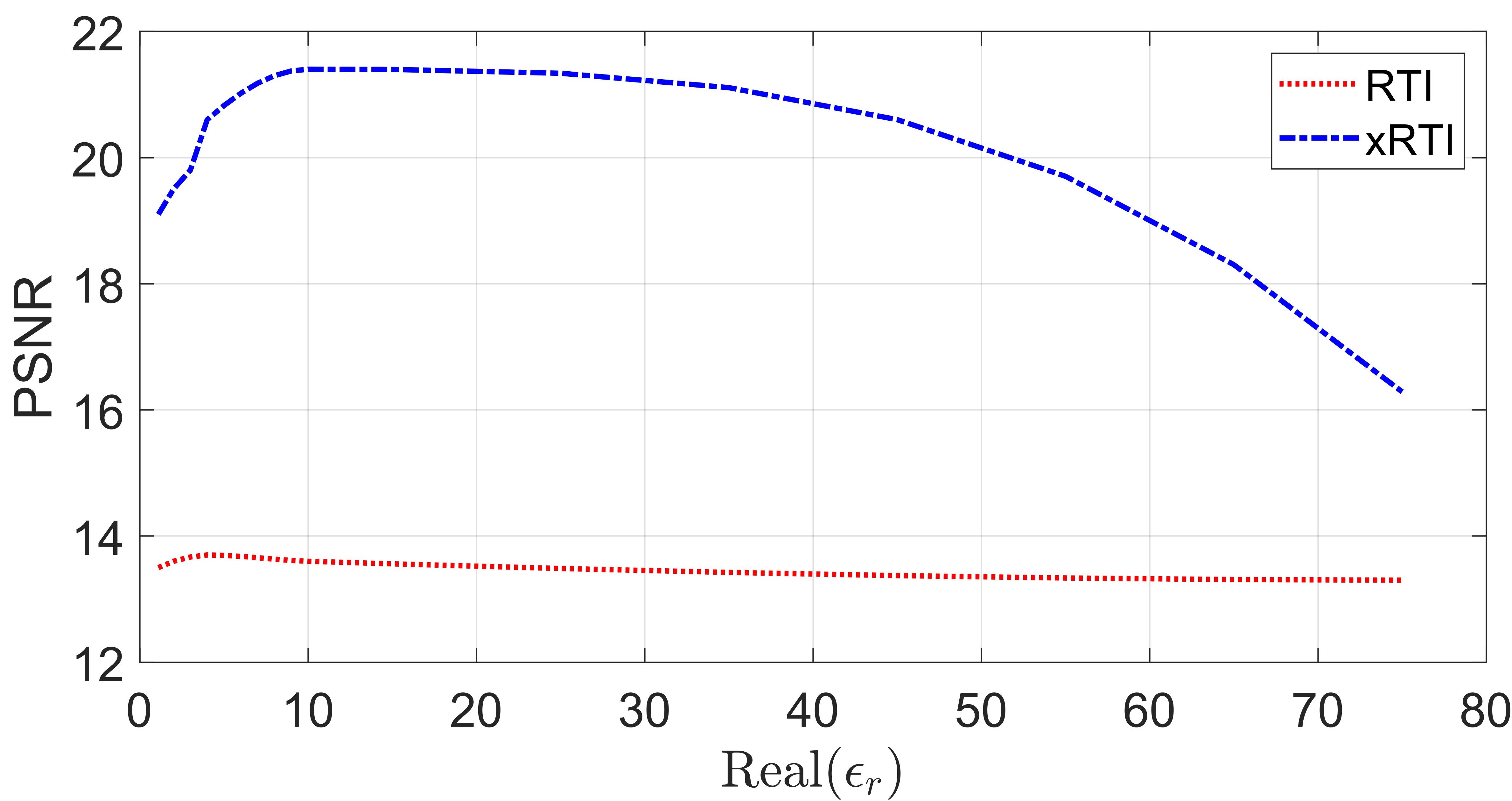}
	\caption{Effect of varying real part of relative permittivity $\epsilon_R$ (with fixed loss tangent $\delta=0.1$) on the PSNR of the reconstructions obtained from xRTI and RTI. Note that $\epsilon_R$ is related to the attenuation parameter $\alpha$ by (\ref{Eq_alpha}).}
	\label{psnr_plot} 
	\vspace{-0.5\baselineskip}		
\end{figure}

From Fig. \ref{Single_scat}(a)-(e), it can be clearly seen that for all the values of permittivity $\epsilon_R$ (or attenuation $\alpha$), xRTI outperforms RTI. 

	
For the weak scattering case $(\epsilon_R=1.1)$ in Fig. \ref{Single_scat}(a), xRTI significantly outperforms RTI. This is because RTI completely relies on the LOS shadowing loss and for a small values of $\epsilon_R=1.1$ (and $\delta=0.1$), the LOS attenuation caused by the object is very small. This small loss along the LOS path does not provide enough information in the RSS measurements for the RTI model to resolve. On the other hand, as explained in the previous sections, xRTI consider both LOS and NLOS paths with proper weighting and this explains their superior performance over RTI. 

For moderate scattering conditions in Fig. \ref{Single_scat}(b)-(c) with $\epsilon_R = 2$ and $4$, xRTI again significantly outperform RTI. 


\begin{figure}[h!]
	\captionsetup[subfigure]{justification=centering}
	\centering
	\begin{subfigure}[b]{1\linewidth}
		\centering
			\adjustbox{minipage=0.4\linewidth}{\subcaption*{\textbf{{\normalsize RTI}} }} \hspace{10mm}
			\adjustbox{minipage=0.4\linewidth}{\subcaption*{\textbf{{\normalsize xRTI}} }} \\
		\includegraphics[width=0.4\linewidth]{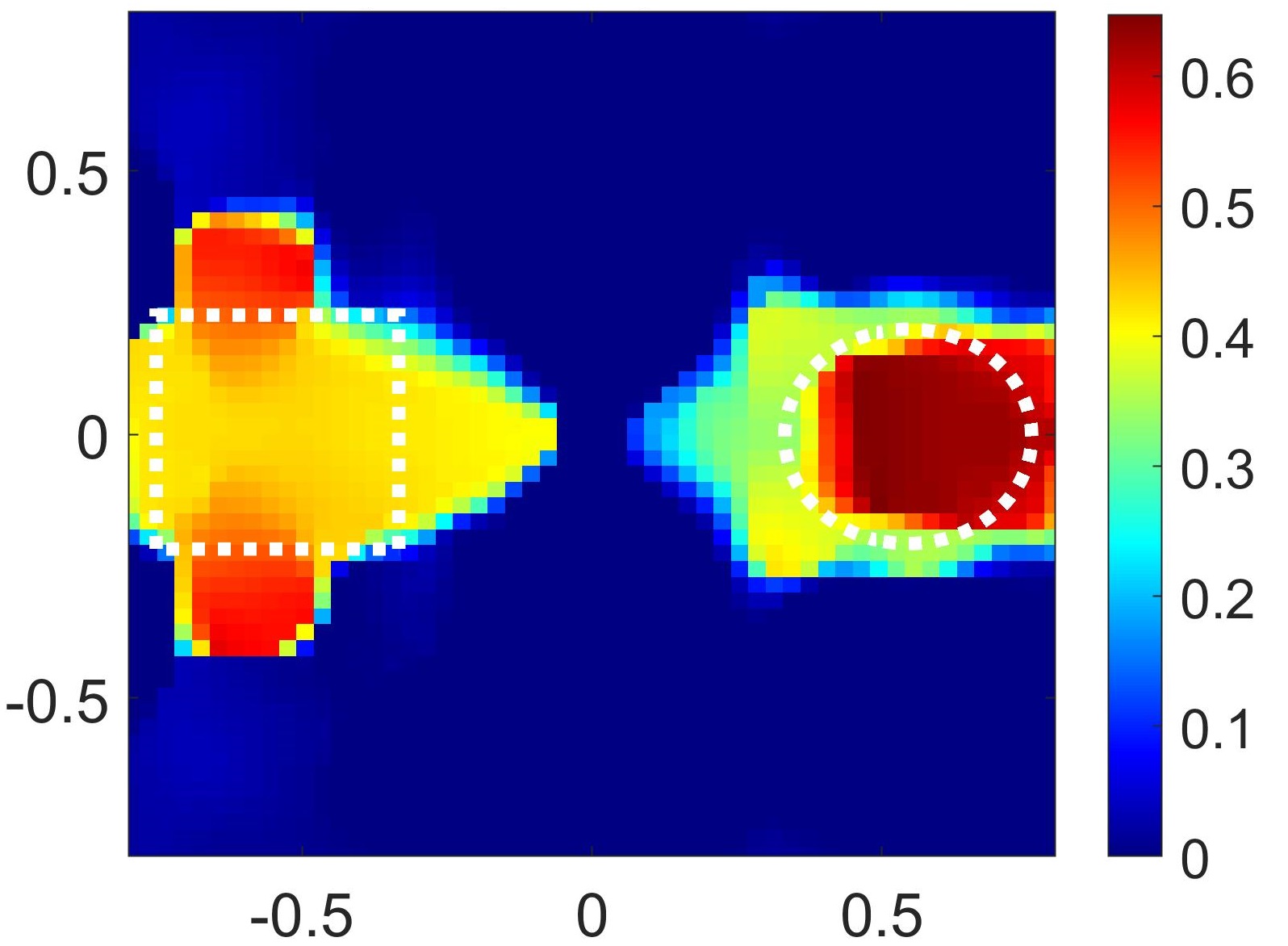} \hspace{10mm}
		\includegraphics[width=0.4\linewidth]{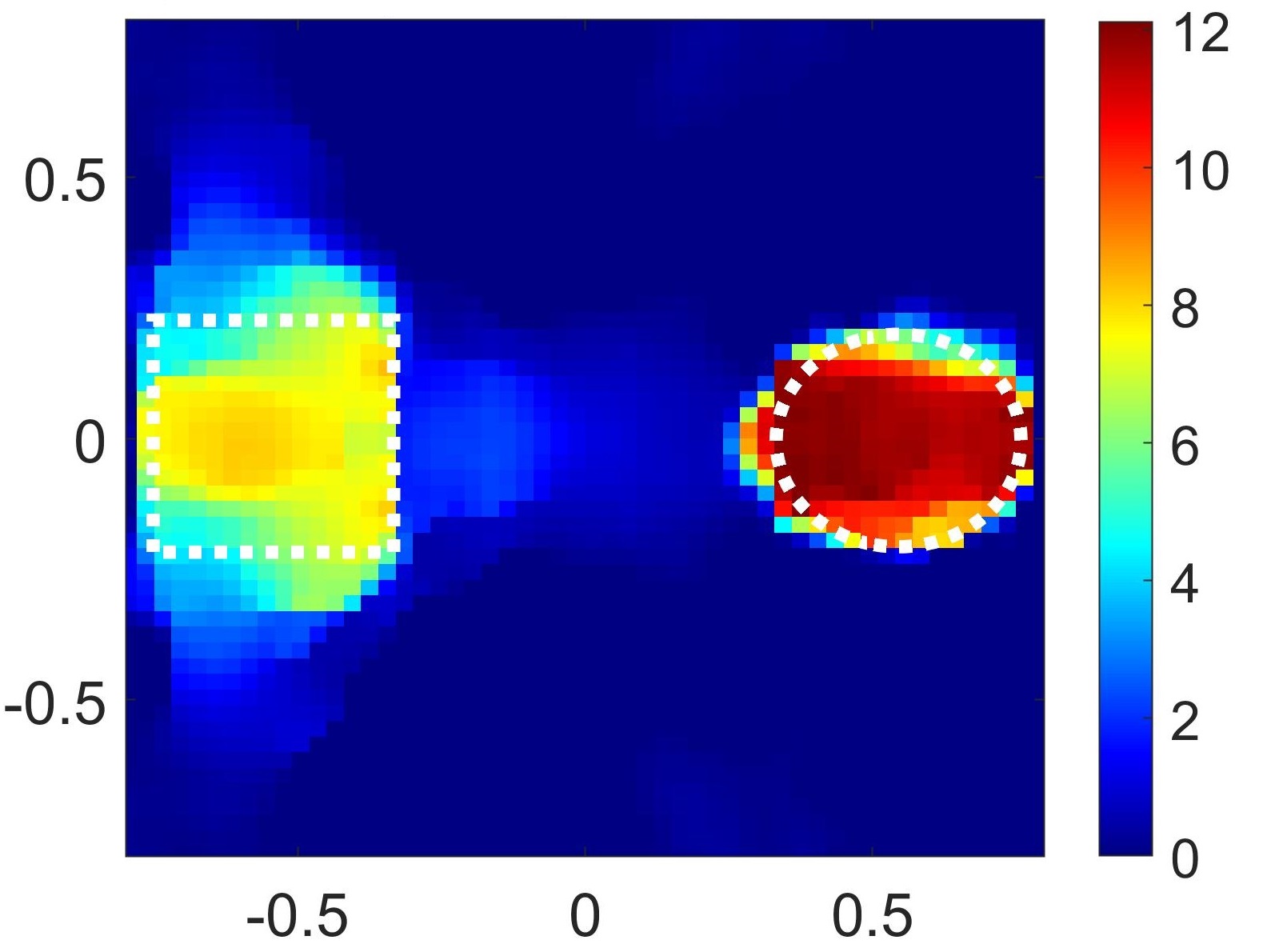}
		\caption{Total measurements used $L=190$ ($M=20$ nodes). The PSNR values for RTI and xRTI are 11.8 dB and 16 dB respectively. The corresponding values of $p$ (in regularization parameter $\Phi=2^p$) are 2 and 4.}
	\end{subfigure}
	\begin{subfigure}[b]{1\linewidth}
		\centering
		\includegraphics[width=0.4\linewidth]{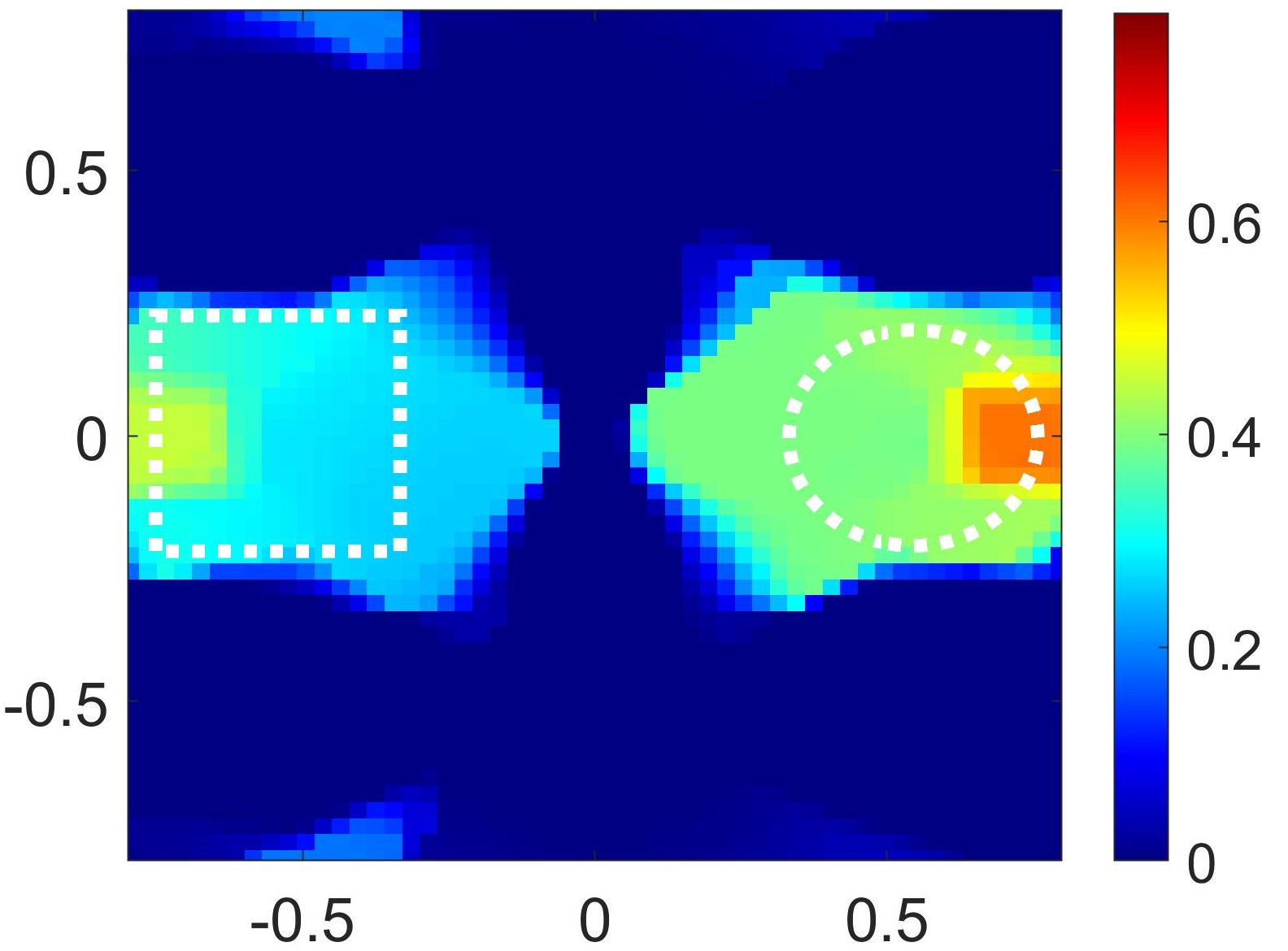} \hspace{10mm}
		\includegraphics[width=0.4\linewidth]{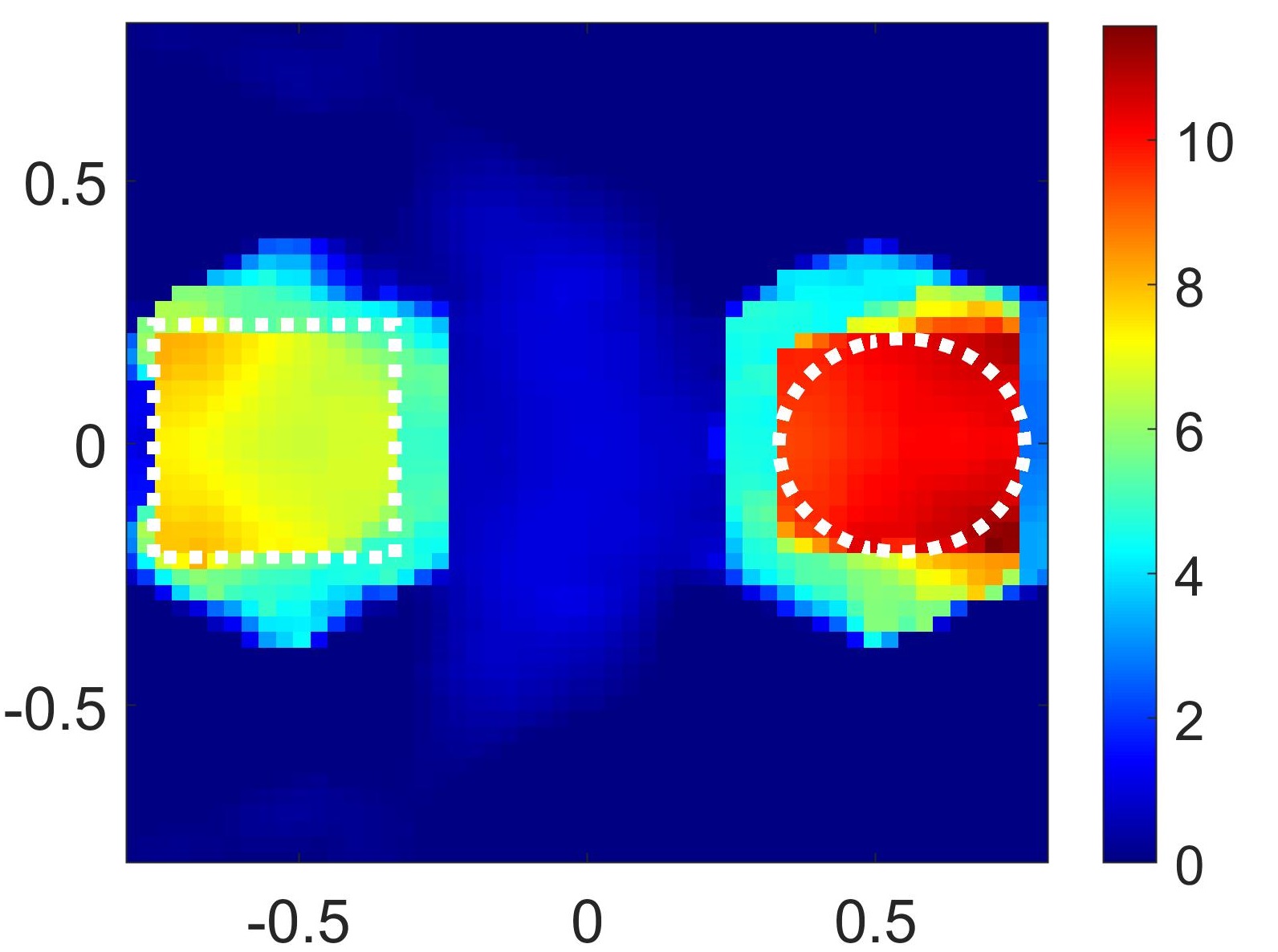}		
		\caption{Total measurements used $L=66$ ($M=12$ nodes). The PSNR values for RTI and xRTI are 11.2 dB and 15 dB respectively. The corresponding values of $p$ (in regularization parameter $\Phi=2^p$) are 2 and 7.}
\end{subfigure}
	\caption{Reconstruction of two objects. The color-scale represent the attenuation parameter $\alpha= {2\pi \delta \sqrt{\epsilon_R}}/{\lambda_0}$. The x and y-axis are in meters.}
	\label{two_dielectric}
\end{figure}

Fig. \ref{Single_scat}(d) and (e) show reconstructions for extremely strong scattering conditions where $\epsilon_R = 10$ and $77$ respectively. It can be seen in Fig. \ref{Single_scat}(d) and (e) that RTI and xRTI provide good shape reconstruction of the scatterer. However, xRTI consistently outperform RTI by a large margin. 

It is also important to note that the PSNR values encapsulate estimation accuracy for both shape and attenuation parameter estimation. Hence, a higher PSNR can be achieved when both shape reconstruction and attenuation parameter estimation is accurate. However, it is important to understand the effect of scattering strength separately on the  shape reconstruction and attenuation parameter estimation (because even shape reconstruction alone can have several applications such as device-free tracking and localization of people inside the indoor region). It can be seen that for both weak and moderate scattering cases (Fig. \ref{Single_scat}(a)-(c)), the reconstruction results show that xRTI outperform RTI not only in terms of shape reconstruction but also in terms of estimating the attenuation parameter (see the colorbar scales of the results). In fact, xRTI provide very accurate attenuation parameter estimation whereas RTI provides highly inaccurate values. RTI cannot provide attenuation parameter estimation because it uses an empirical model for attenuation as explained in section \ref{Sec_Analysis}. For extremely strong scattering cases Fig. \ref{Single_scat}(d)-(e), however, xRTI also fails to accurately estimate the attenuation parameter (but the shape reconstruction remains accurate).

In Fig. \ref{psnr_plot}, we evaluate the PSNR performance by varying the real part of permittivity $\epsilon_R$ over a large range between $1.1$ to $75$ while keeping the loss tangent fixed at ($\delta=0.1$). Similar to the results in Fig \ref{Single_scat}, it can be seen that for all values of permittivity, xRTI significantly outperform RTI. Also, we can see that for higher permittivity, the performance of xRTI decreases. This decline in performance is due to inaccurate attenuation parameter estimation, (however as evident from Fig \ref{Single_scat}, the shape reconstruction performance of xRTI remains excellent even for higher permittivity values). 

To summarize, we can conclude from Fig. \ref{Single_scat} and Fig. \ref{psnr_plot} that xRTI significantly outperforms RTI. 

It is also important to note that if the target objects exhibit low-loss, have piecewise homogeneous distribution of permittivity, and are electrically large, both xRTI and RTI provide good shape reconstruction up to $\epsilon_R\le 77$, whereas, the validity range of state-of-the-art phaseless nonlinear methods does not currently extend to $\epsilon_R\le 77$   \cite{dubeyTGRS, Xudongchen, chen2010subspace, chen2018computational}. The competitive performance of xRTI can be justified by the derivations described here and also in the description of xPRA-LM \cite{dubeyTGRS}. However, for RTI it is counter-intuitive that its shape reconstruction performance improves as the permittivity becomes higher. The expectation is that it should become worse as for formal inverse scattering approaches \cite{chen2018computational, Xudongchen, chen2010subspace}. One reason for RTI's shape reconstruction performance at high permittivity values is that the technique is based on LOS attenuation (and neglects diffraction and scattering) and performs better when this assumption is more closely met. In Fig \ref{Single_scat}, even though the loss tangent is small $\delta = 0.1$, as we increase the real part of permittivity $(\epsilon_R)$, the imaginary part of permittivity also increases $(\epsilon_I = 0.1\epsilon_R)$. For objects that are large in size and have high $\epsilon_I$, the attenuation caused by the object on the incident signal becomes large and creates a shadow region similar to LOS propagation. Hence, RTI provides better shape reconstruction in this scenario.

Fig. \ref{two_dielectric} provides reconstruction results for the two object profile shown in Fig. \ref{Single_Scat_GT}(b). The reconstruction results using RTI and xRTI with $M=20$ nodes (or $L=190$ measurements) are shown in Fig. \ref{two_dielectric}(a). These reconstructions show that xRTI outperform RTI in handling the multiple scattering between two strong scatterers. 
In general, the resolvability of two objects depends on how strong the multiple scattering between the two objects is. Therefore, the resolvable distance between two objects depends on multiple factors such as the size, shape, and permittivity of the objects. Hence, there is no unique answer to the value of the resolvable distance between the objects. However, based on our numerical analysis, the reconstruction quality deteriorates as object separation becomes comparable or equal to the wavelength ($\lambda_0$). In particular, as the separation becomes equal to $\lambda_0$, the objects are no longer resolvable.

Fig. \ref{two_dielectric}(b) show reconstruction results with reduced number of Wi-Fi nodes $M=12$. For these reconstructions with $12$ nodes, now the number of measurements becomes $L=66$ compared to $L=190$ for 20 nodes configuration in previous simulation examples. Even with the smaller number of measurements, xRTI are able to provide better performance whereas, RTI fails to provide acceptable reconstruction.


\begin{figure}[h!]
	\centering
	\begin{subfigure}[t]{0.26\textwidth}
		\centering
		\includegraphics[width=\textwidth]{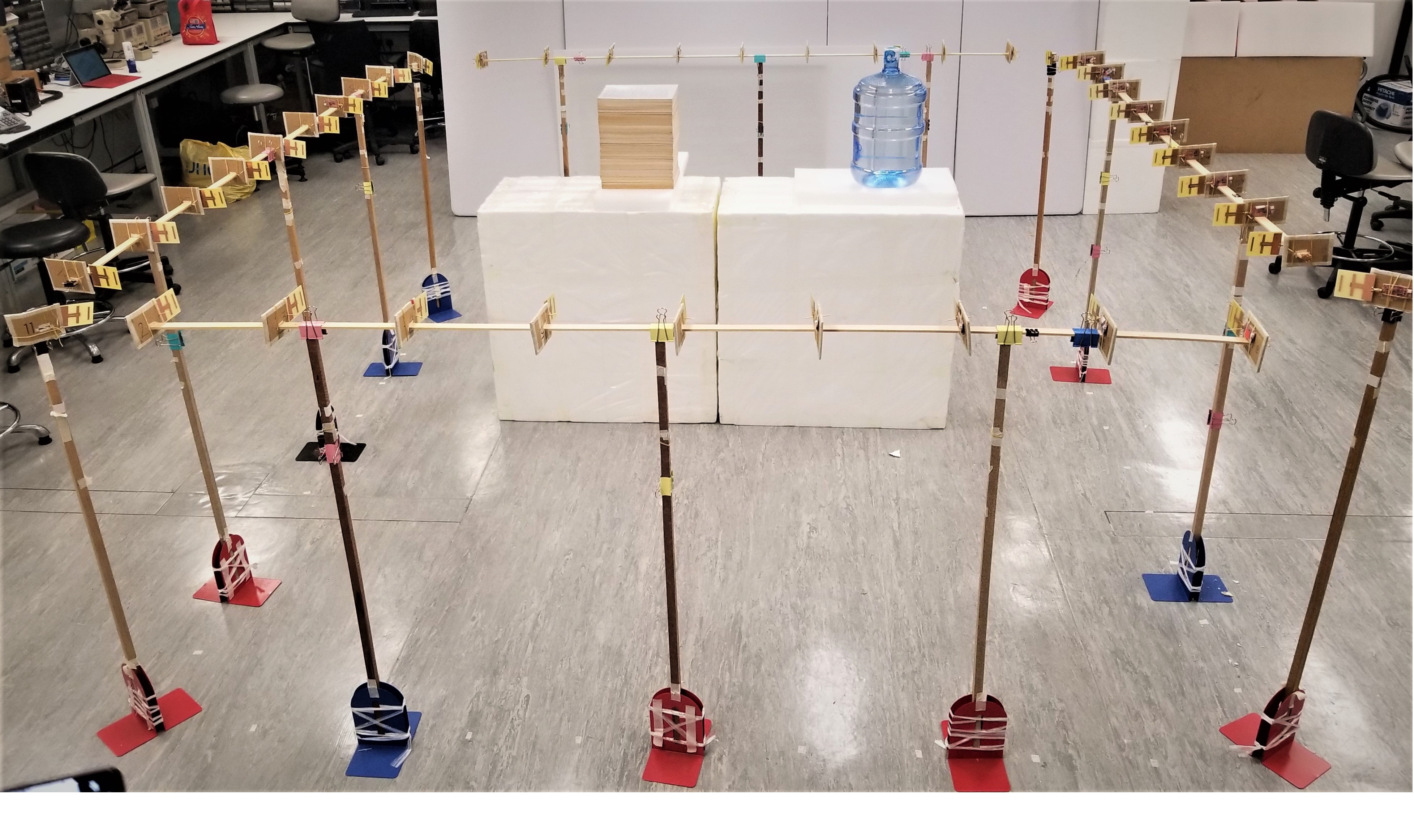}
		\subcaption{Experiment Setup}
	\end{subfigure}
	\begin{subfigure}[t]{0.215\textwidth}
		\centering
		\includegraphics[width=\textwidth]{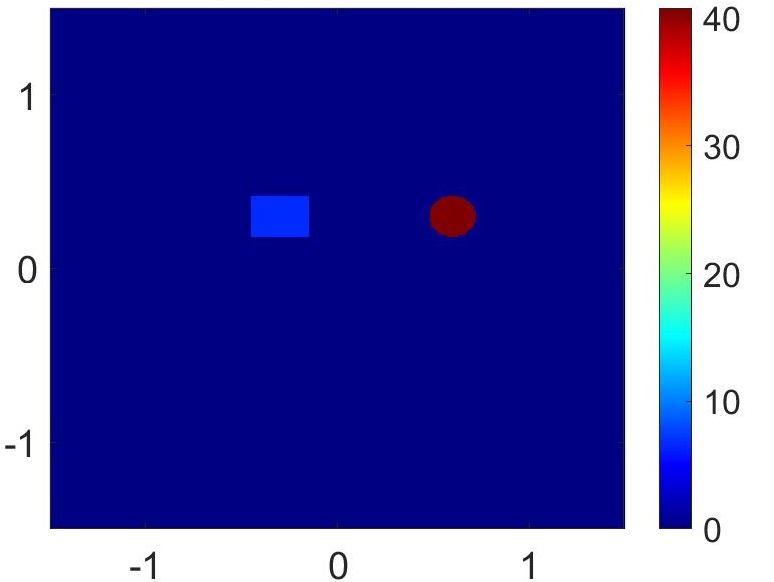}
		\subcaption{Ground truth attenuation profile}
	\end{subfigure}
	\caption{Experimental setup. (a) Experimental setup with $3\times 3$ m$^2$ DOI with Wi-Fi transceiver nodes placed around the boundaries. The DOI has two scattering objects, including a stack of books with $\epsilon_r=3.4+0.25j$ and a water container with $\epsilon_r=77+7j$, (b) shows 2D ground truth attenuation profile $\alpha= {2\pi \delta \sqrt{\epsilon_R}}/{\lambda_0}$.}
	\label{ExpGT}
\end{figure}

\begin{figure}[h!]
	\captionsetup[subfigure]{justification=centering}
	\centering
	\begin{subfigure}[b]{1\linewidth}
		\centering
		\adjustbox{minipage=0.4\linewidth}{\subcaption*{\textbf{{\normalsize RTI}} }} \hspace{10mm}
		\adjustbox{minipage=0.4\linewidth}{\subcaption*{\textbf{{\normalsize xRTI}} }} \\
		\includegraphics[width=0.4\linewidth]{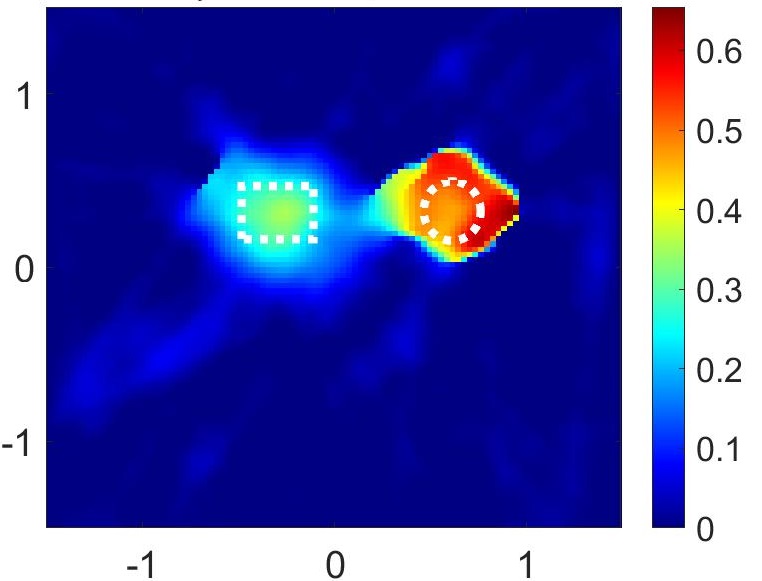} \hspace{10mm}
		\includegraphics[width=0.4\linewidth]{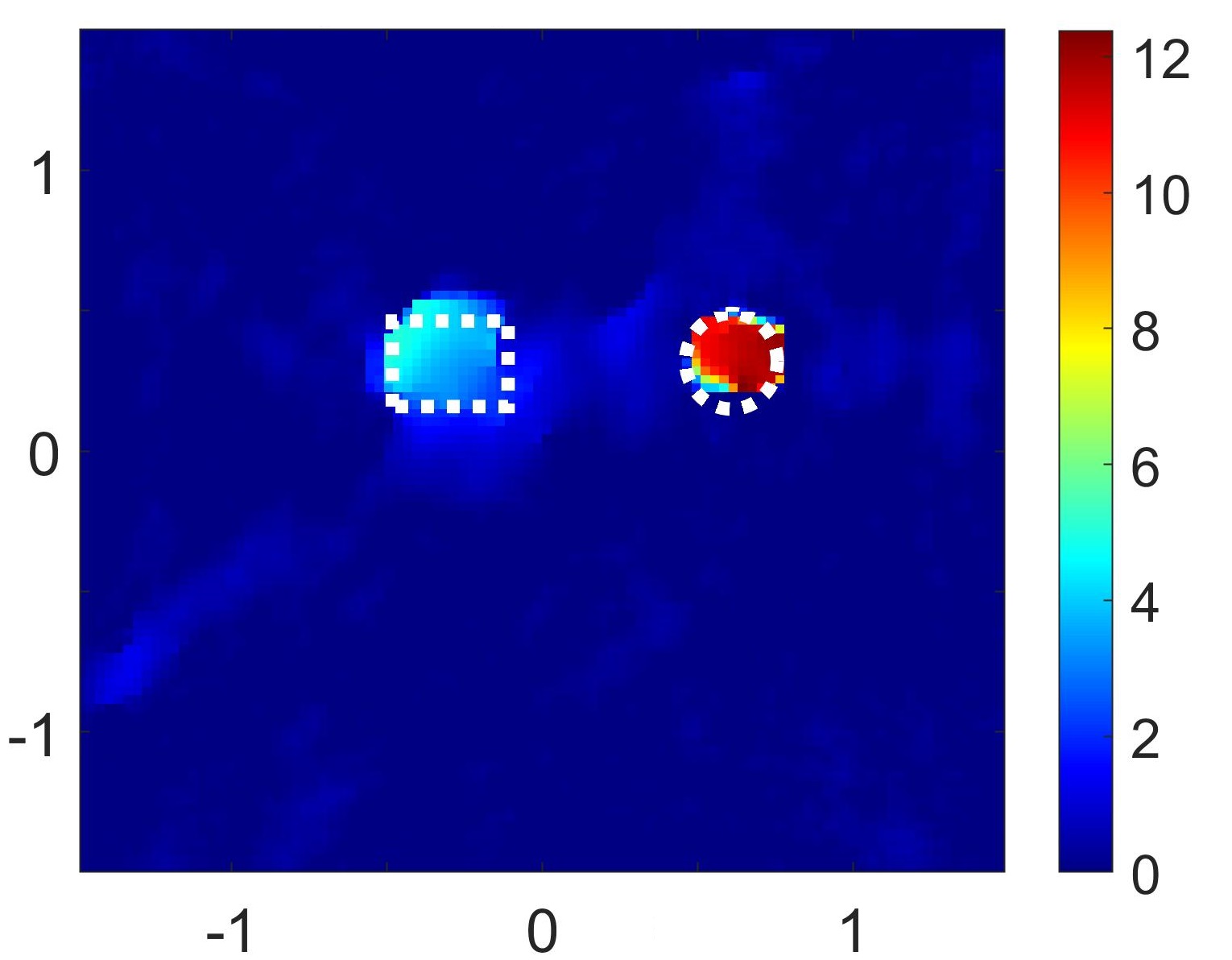}
	\end{subfigure}
	\caption{Experimental results for ground truth profile shown in Fig. \ref{ExpGT}. The PSNR values for RTI and xRTI are 17 dB and 19.8 dB respectively. The corresponding values of $p$ (in regularization parameter $\Phi=2^p$) are 2 and 5. The x and y-axis are in meters.}
	\label{Exp}
\end{figure}

\subsection{Experimental Results}

Fig. \ref{problemsetup}(b) and Fig. \ref{ExpGT} show our experimental setup. They show a $3\times 3$ m$^2$ imaging region located in the room 3125A at the Hong Kong University of Science and Technology (HKUST). The goal is to image the 2D DOI in this 3D region.

We have utilized an object configuration similar in form to that in Fig. \ref{Single_Scat_GT}b. In particular in Fig. \ref{ExpGT}(a) we show the test scatterers including a stack of books and a circular plastic container filled with water placed upon a Styrofoam platform. The book stack has a rectangular cross section of $30\times 21$ cm$^2$ and the container of water has a circular cross section with diameter $26$ cm. The book stack acts as a strong scatterer with relative permittivity $\epsilon_r= 3.4 + j 0.25$ (measured using cavity resonator) and the water container acts as a very strong scatterer with relative permittivity $\epsilon_r= 77 + j 7$ at 2.4 GHz \cite{4562803, ahmad2014partially, Productnote}. Fig. \ref{ExpGT}(b) shows the ground truth attenuation parameter profile representing the 2D cross section of the experimental setup shown in Fig. \ref{ExpGT}(a).

The Wi-Fi transceiver nodes are placed at the boundary of this DOI. Each node consists of a SparkFun ESP32 Thing board \cite{ESP3200} consisting of an integrated 802.11bgn Wi-Fi transceiver operating at 2.4 GHz. The inbuilt omni-directional antenna of the ESP32 boards are replaced by a Yagi antenna of 6.6 dBi. The ESP32 boards are located at a height of $d_h=1.2$ m from the floor using a wooden stand (see Fig. \ref{problemsetup}(b) and Fig. \ref{ExpGT}(a)). The Yagi antennas on the boards are oriented such that the xz-plane in Fig. \ref{ExpGT}(a) lies in the 2D DOI plane and yz-plane (center dipole element of the antenna) is normal to 2D DOI plane and hence matches our TM (vertical polarization) simulations and formulations. Every transceiver can be assumed to alternate between access point (AP) and station (STA) mode so that the Wi-Fi beacon signal can be utilized to obtain the RSSI for each link.

At height $d_h = 1.2$ m, the 2D cross sectional slice of 3D DOI in Fig. \ref{ExpGT}(a) matches our simulation setup where DOI dimensions and locations of nodes are exactly the same as in Fig. \ref{ExpGT}(a). There are differences between the simulations and experiments in which the simulation setup is an ideal 2D environment whereas the experimental setup is a 3D environment with 2D DOI (at height 1.2 m from floor). Therefore, the experimental measurements contain distortions due to multipath reflections from the ceiling, floor and walls and other background clutter outside the DOI. Furthermore, experimental data also contains errors and noise which is absent in the simulations. We use Yagi antennas to reduce the effect of these 3D multipath reflections from ceiling and floor. However, the multiple scattering within the 2D domain and multipath distortions due to background clutter from walls and other objects in that plane still remain and must be handled by RTI and xRTI. The key technique which can handle these distortions and errors in the experimental setup is the temporal background subtraction framework which is incorporated in both the techniques, RTI and xRTI (as explained in Section \ref{Sec_Analysis}). The temporal background subtraction can be implemented by collecting RSS data in the presence and in the absence of the target objects and using these two RSS measurements to estimate change in RSS values $\Delta \mybar{\textbf{P}}$ using RTI (\ref{Eq_ConRTI_LS}) and xRTI (\ref{Eq_ConxRTI}) models.


The reconstruction results for Fig. \ref{ExpGT} are shown in Fig. \ref{Exp}. It can be seen that the reconstructions using xRTI are significantly better than the reconstruction obtained using RTI  (PSNR of 19.8 dB compared to 17 dB). Such an experimental demonstration can only be achieved using techniques such as RTI and xRTI due to their linear phaseless form which enables us to include temporal background subtraction and this is not possible with state-of-the-art non-linear and deep learning techniques. 

\section{Conclusion and Future Work}
In this paper we reconcile an empirically motivated RF imaging technique, RTI, with a formal inverse scattering approach based on xPRA-LM. We derive and highlight key mathematical differences and similarities between RTI and xPRA-LM. In particular we show that along the LOS path, the empirically estimated attenuation model of RTI has a physics based justification that can be related to xPRA-LM. This provides justification for the good performance of RTI in extremely strong scattering and low-loss conditions where formal inverse scattering techniques are often challenged. However RTI deviates significantly from xPRA-LM in other aspects and therefore modifications can be included in RTI to improve its performance. 
As such we also propose straightforward enhancements to RTI to significantly improve its performance and denote this enhanced RTI as xRTI. This enhancement involves replacing the empirical attenuation model of RTI with a xPRA-LM based attenuation model. With simulation and experimental results in a real indoor environment with off-the-shelf 2.4 GHz Wi-Fi devices, we show that xRTI outperforms RTI. 

Future work can include making RTI and xRTI practically feasible. Both RTI and xPRA-LM are challenged by the requirement for the large number of transceivers required. To address this limitation, methods such as frequency \cite{9967760} and pattern diversity can be used \cite{10241308} along with data-driven regularization \cite{Unrolled}. In addition, it would also be useful to reconcile the differences between radar and inverse scattering based techniques so the features of both can be achieved. For example radar only requires one sensing location while inverse scattering techniques can obtain estimates of the materials properties of objects.

\bibliographystyle{IEEEtran}
\bibliography{main}

\begin{thebibliography}{10}
\providecommand{\url}[1]{#1}
\csname url@samestyle\endcsname
\providecommand{\newblock}{\relax}
\providecommand{\bibinfo}[2]{#2}
\providecommand{\BIBentrySTDinterwordspacing}{\spaceskip=0pt\relax}
\providecommand{\BIBentryALTinterwordstretchfactor}{4}
\providecommand{\BIBentryALTinterwordspacing}{\spaceskip=\fontdimen2\font plus
\BIBentryALTinterwordstretchfactor\fontdimen3\font minus
  \fontdimen4\font\relax}
\providecommand{\BIBforeignlanguage}[2]{{%
\expandafter\ifx\csname l@#1\endcsname\relax
\typeout{** WARNING: IEEEtran.bst: No hyphenation pattern has been}%
\typeout{** loaded for the language `#1'. Using the pattern for}%
\typeout{** the default language instead.}%
\else
\language=\csname l@#1\endcsname
\fi
#2}}
\providecommand{\BIBdecl}{\relax}
\BIBdecl

\bibitem{depatla2015x}
S.~Depatla, L.~Buckland, and Y.~Mostofi, ``X-ray vision with only {WiFi} power
  measurements using {R}ytov wave models,'' \emph{IEEE Transactions on
  Vehicular Technology}, vol.~64, no.~4, pp. 1376--1387, 2015.

\bibitem{6881182}
F.~{Guidi}, A.~{Guerra}, and D.~{Dardari}, ``Millimeter-wave massive arrays for
  indoor {SLAM},'' in \emph{2014 IEEE International Conference on
  Communications Workshops (ICC)}, 2014, pp. 114--120.

\bibitem{1545232}
E.~{Jose} and M.~D. {Adams}, ``An augmented state {SLAM} formulation for
  multiple line-of-sight features with millimetre wave radar,'' in \emph{2005
  IEEE/RSJ International Conference on Intelligent Robots and Systems}, 2005,
  pp. 3087--3092.

\bibitem{Pastina2015}
D.~Pastina, F.~Colone, T.~Martelli, and P.~Falcone, ``Parasitic exploitation of
  {WiFi} signals for indoor radar surveillance,'' \emph{IEEE Transactions on
  Vehicular Technology}, vol.~64, no.~4, pp. 1401--1415, 2015.

\bibitem{8396311}
L.~{Zhang}, Q.~{Gao}, X.~{Ma}, J.~{Wang}, T.~{Yang}, and H.~{Wang}, ``{DeFi}:
  Robust training-free device-free wireless localization with {WiFi},''
  \emph{IEEE Transactions on Vehicular Technology}, vol.~67, no.~9, pp.
  8822--8831, 2018.

\bibitem{chetty2011through}
K.~Chetty, G.~E. Smith, and K.~Woodbridge, ``Through-the-wall sensing of
  personnel using passive bistatic {WiFi} radar at standoff distances,''
  \emph{IEEE Transactions on Geoscience and Remote Sensing}, vol.~50, no.~4,
  pp. 1218--1226, 2011.

\bibitem{adib2013see}
F.~Adib and D.~Katabi, ``See through walls with {WiFi}!'' in \emph{Proceedings
  of the ACM SIGCOMM 2013 Conference on SIGCOMM}, 2013, pp. 75--86.

\bibitem{adib2015multi}
F.~Adib, Z.~Kabelac, and D.~Katabi, ``Multi-person localization via {RF} body
  reflections,'' in \emph{12th $\{$USENIX$\}$ Symposium on Networked Systems
  Design and Implementation ($\{$NSDI$\}$ 15)}, 2015, pp. 279--292.

\bibitem{Guo2017}
X.~{Guo} and N.~{Ansari}, ``Localization by fusing a group of fingerprints via
  multiple antennas in indoor environment,'' \emph{IEEE Transactions on
  Vehicular Technology}, vol.~66, no.~11, pp. 9904--9915, 2017.

\bibitem{WangFIngerprint2017}
X.~{Wang}, L.~{Gao}, S.~{Mao}, and S.~{Pandey}, ``{CSI}-based fingerprinting
  for indoor localization: A deep learning approach,'' \emph{IEEE Transactions
  on Vehicular Technology}, vol.~66, no.~1, pp. 763--776, 2017.

\bibitem{shi2018accurate}
S.~Shi, S.~Sigg, L.~Chen, and Y.~Ji, ``Accurate location tracking from
  {CSI}-based passive device-free probabilistic fingerprinting,'' \emph{IEEE
  Transactions on Vehicular Technology}, vol.~67, no.~6, pp. 5217--5230, 2018.

\bibitem{sun2018single}
X.~Sun, X.~Gao, G.~Y. Li, and W.~Han, ``Single-site localization based on a new
  type of fingerprint for massive {MIMO-OFDM} systems,'' \emph{IEEE
  Transactions on Vehicular Technology}, vol.~67, no.~7, pp. 6134--6145, 2018.

\bibitem{liu2020massive}
B.~Liu, A.~P. Guevara, S.~De~Bast, Q.~Wang, and S.~Pollin, ``Massive {MIMO}
  indoor localization with 64-antenna uniform linear array,'' in \emph{2020
  IEEE 91st Vehicular Technology Conference (VTC2020-Spring)}.\hskip 1em plus
  0.5em minus 0.4em\relax IEEE, 2020, pp. 1--5.

\bibitem{zhang2015generalized}
W.~Zhang and A.~Hoorfar, ``A generalized approach for {SAR} and {MIMO} radar
  imaging of building interior targets with compressive sensing,'' \emph{IEEE
  Antennas and Wireless Propagation Letters}, vol.~14, pp. 1052--1055, 2015.

\bibitem{IPIN}
A.~{Zayets} and E.~{Steinbach}, ``Robust {WiFi}-based indoor localization using
  multipath component analysis,'' in \emph{2017 International Conference on
  Indoor Positioning and Indoor Navigation (IPIN)}, 2017, pp. 1--8.

\bibitem{7426565}
K.~{Witrisal}, P.~{Meissner}, E.~{Leitinger}, Y.~{Shen}, C.~{Gustafson},
  F.~{Tufvesson}, K.~{Haneda}, D.~{Dardari}, A.~F. {Molisch}, A.~{Conti}, and
  M.~Z. {Win}, ``High-accuracy localization for assisted living: {5G} systems
  will turn multipath channels from foe to friend,'' \emph{IEEE Signal
  Processing Magazine}, vol.~33, no.~2, pp. 59--70, 2016.

\bibitem{Patwari2017}
O.~{Kaltiokallio}, R.~{Jäntti}, and N.~{Patwari}, ``{ARTI}: An adaptive radio
  tomographic imaging system,'' \emph{IEEE Transactions on Vehicular
  Technology}, vol.~66, no.~8, pp. 7302--7316, 2017.

\bibitem{Patwari2015}
B.~Wei, A.~Varshney, N.~Patwari, W.~Hu, T.~Voigt, and C.~T. Chou, ``{dRTI}:
  Directional radio tomographic imaging,'' in \emph{Proceedings of the 14th
  International Conference on Information Processing in Sensor Networks
  (IPSN)}, 2015, pp. 166--177.

\bibitem{Patwari2014}
O.~{Kaltiokallio}, M.~{Bocca}, and N.~{Patwari}, ``A fade level-based spatial
  model for radio tomographic imaging,'' \emph{IEEE Transactions on Mobile
  Computing}, vol.~13, no.~6, pp. 1159--1172, 2014.

\bibitem{Patwari2010}
J.~{Wilson} and N.~{Patwari}, ``Radio tomographic imaging with wireless
  networks,'' \emph{IEEE Transactions on Mobile Computing}, vol.~9, no.~5, pp.
  621--632, 2010.

\bibitem{1Patwari2013}
Y.~{Zhao}, N.~{Patwari}, J.~M. {Phillips}, and S.~{Venkatasubramanian}, ``Radio
  tomographic imaging and tracking of stationary and moving people via kernel
  distance,'' in \emph{2013 ACM/IEEE International Conference on Information
  Processing in Sensor Networks (IPSN)}, 2013, pp. 229--240.

\bibitem{Savazzi2014}
S.~{Savazzi}, M.~{Nicoli}, F.~{Carminati}, and M.~{Riva}, ``A {Bayesian}
  approach to device-free localization: Modeling and experimental assessment,''
  \emph{IEEE Journal of Selected Topics in Signal Processing}, vol.~8, no.~1,
  pp. 16--29, 2014.

\bibitem{WangRTI32016}
J.~{Wang}, Q.~{Gao}, M.~{Pan}, X.~{Zhang}, Y.~{Yu}, and H.~{Wang}, ``Toward
  accurate device-free wireless localization with a saddle surface model,''
  \emph{IEEE Transactions on Vehicular Technology}, vol.~65, no.~8, pp.
  6665--6677, 2016.

\bibitem{WangRTI22017}
J.~{Wang}, X.~{Zhang}, Q.~{Gao}, X.~{Ma}, X.~{Feng}, and H.~{Wang},
  ``Device-free simultaneous wireless localization and activity recognition
  with wavelet feature,'' \emph{IEEE Transactions on Vehicular Technology},
  vol.~66, no.~2, pp. 1659--1669, 2017.

\bibitem{WangRTI2016}
Q.~{Wang}, H.~{Yiğitler}, R.~{Jäntti}, and X.~{Huang}, ``Localizing multiple
  objects using radio tomographic imaging technology,'' \emph{IEEE Transactions
  on Vehicular Technology}, vol.~65, no.~5, pp. 3641--3656, 2016.

\bibitem{TomicRTI2015}
S.~{Tomic}, M.~{Beko}, and R.~{Dinis}, ``{RSS}-based localization in wireless
  sensor networks using convex relaxation: Noncooperative and cooperative
  schemes,'' \emph{IEEE Transactions on Vehicular Technology}, vol.~64, no.~5,
  pp. 2037--2050, 2015.

\bibitem{chen2018computational}
X.~Chen, \emph{Computational methods for electromagnetic inverse
  scattering}.\hskip 1em plus 0.5em minus 0.4em\relax Wiley Online Library,
  2018.

\bibitem{bates1991manageable}
R.~Bates, V.~Smith, and R.~Murch, ``Manageable multidimensional inverse
  scattering theory,'' \emph{Physics Reports}, vol. 201, no.~4, pp. 185--277,
  1991.

\bibitem{Xudongchen}
K.~Xu, L.~Wu, X.~Ye, and X.~Chen, ``Deep learning-based inversion methods for
  solving inverse scattering problems with phaseless data,'' \emph{IEEE
  Transactions on Antennas and Propagation}, vol.~68, no.~11, pp. 7457--7470,
  2020.

\bibitem{chen2010subspace}
L.~Pan, Y.~Zhong, X.~Chen, and S.~P. Yeo, ``Subspace-based optimization method
  for inverse scattering problems utilizing phaseless data,'' \emph{IEEE
  Transactions on Geoscience and Remote Sensing}, vol.~49, no.~3, pp. 981--987,
  2010.

\bibitem{dubeyTGRS}
A.~Dubey, S.~Deshmukh, L.~Pan, X.~Chen, and R.~Murch, ``A phaseless extended
  rytov approximation for strongly scattering low-loss media and its
  application to indoor imaging,'' \emph{IEEE Transactions on Geoscience and
  Remote Sensing}, vol.~60, pp. 1--17, 2022.

\bibitem{komarov2005permittivity}
V.~Komarov, S.~Wang, and J.~Tang, ``Permittivity and measurements,''
  \emph{Encyclopedia of RF and Microwave Engineering}, 2005.

\bibitem{Productnote}
K.~C. Yaw, ``Measurement of dielectric material properties,'' \emph{Application
  Note. Rohde \& Schwarz}, pp. 1--35, 2012.

\bibitem{4562803}
Y.~Pinhasi, A.~Yahalom, and S.~Petnev, ``Propagation of ultra wide-band signals
  in lossy dispersive media,'' in \emph{2008 IEEE International Conference on
  Microwaves, Communications, Antennas and Electronic Systems}, 2008, pp.
  1--10.

\bibitem{10241308}
D.~Ma, A.~Dubey, Z.~Xu, S.~Shen, Q.~Zhang, and R.~Murch, ``Reducing the number
  of measurement nodes in rf imaging using antenna-pattern diversity with an
  extended rytov approximation,'' \emph{IEEE Transactions on Antennas and
  Propagation}, vol.~71, no.~11, pp. 8881--8893, 2023.

\bibitem{9967760}
D.~Ma, Y.~Zhang, A.~Dubey, S.~Deshmukh, S.~Shen, Q.~Zhang, and R.~Murch,
  ``Millimeter-wave 3-d imaging using leaky-wave antennas and an extended rytov
  approximation in a frequency-diverse mimo system,'' \emph{IEEE Transactions
  on Microwave Theory and Techniques}, vol.~71, no.~4, pp. 1809--1825, 2023.

\bibitem{hecht2012optics}
E.~Hecht, \emph{Optics}.\hskip 1em plus 0.5em minus 0.4em\relax Pearson
  Education, 2012.

\bibitem{bornwolf}
M.~Born and E.~Wolf, \emph{Principles of optics: electromagnetic theory of
  propagation, interference and diffraction of light}.\hskip 1em plus 0.5em
  minus 0.4em\relax Elsevier, 2013.

\bibitem{DubeyTAP}
A.~Dubey, X.~Chen, and R.~Murch, ``A new correction to the rytov approximation
  for strongly scattering lossy media,'' \emph{IEEE Transactions on Antennas
  and Propagation}, vol.~70, no.~11, pp. 10\,851--10\,864, 2022.

\bibitem{wu2003wave}
R.-S. Wu, ``Wave propagation, scattering and imaging using dual-domain one-way
  and one-return propagators,'' \emph{Pure and Applied Geophysics}, vol. 160,
  no.~3, pp. 509--539, 2003.

\bibitem{li2010efficient}
C.~Li, \emph{An efficient algorithm for total variation regularization with
  applications to the single pixel camera and compressive sensing}.\hskip 1em
  plus 0.5em minus 0.4em\relax Rice University, 2010.

\bibitem{li2009user}
C.~Li, W.~Yin, and Y.~Zhang, ``User’s guide for {TVAL3}: {TV} minimization by
  augmented {Lagrangian} and alternating direction algorithms,'' \emph{CAAM
  Report}, vol.~20, no. 46-47, p.~4, 2009.

\bibitem{Unrolled}
S.~Deshmukh, A.~Dubey, and R.~Murch, ``Unrolled optimization with deep
  learning-based priors for phaseless inverse scattering problems,'' \emph{IEEE
  Transactions on Geoscience and Remote Sensing}, vol.~60, pp. 1--14, 2022.

\bibitem{DeshmukhTAP}
S.~Deshmukh, A.~Dubey, D.~Ma, Q.~Chen, and R.~Murch, ``Physics assisted deep
  learning for indoor imaging using phaseless wi-fi measurements,'' \emph{IEEE
  Transactions on Antennas and Propagation}, vol.~70, no.~10, pp. 9716--9731,
  2022.

\bibitem{ahmad2014partially}
F.~Ahmad, M.~G. Amin, and T.~Dogaru, ``Partially sparse imaging of stationary
  indoor scenes,'' \emph{EURASIP Journal on Advances in Signal Processing},
  vol. 2014, no.~1, pp. 1--15, 2014.

\bibitem{ESP3200}
\BIBentryALTinterwordspacing
S.~Espressif, ``{ESP32} datasheet,'' 2016. [Online]. Available:
  \url{https://www.sparkfun.com/products/13907}
\BIBentrySTDinterwordspacing

\end{thebibliography}
%
\end{document}